\documentclass[%
reprint,
superscriptaddress,
tbtags,
amsmath,amssymb,
aps,
prl,
floatfix,
]{revtex4-1}

\usepackage{graphicx}
\usepackage{bm}
\usepackage{simpler-wick}
\usepackage[colorlinks=true, citecolor=blue]{hyperref}
\usepackage{xcolor}
\usepackage{enumitem}
\usepackage{float}
\usepackage{booktabs}
\allowdisplaybreaks

\AtBeginDocument{%
  \heavyrulewidth=.1em
  \lightrulewidth=.05em
  \cmidrulewidth=.03em
  \belowrulesep=.65ex
  \belowbottomsep=0pt
  \aboverulesep=.4ex
  \abovetopsep=0pt
  \cmidrulesep=\doublerulesep
  \cmidrulekern=.5em
  \defaultaddspace=.5em
}
\newcommand{\rss}{\rho_{\rm{ss}}^{}}
\newcommand{\he}{\mathcal{H}_E}
\newcommand{\hi}{\mathcal{H}_I}
\newcommand{\jih}{\mathcal{J}_{I_h}}
\newcommand{\jic}{\mathcal{J}_{I_c}}
\newcommand{\jeh}{\mathcal{J}_{E_h}}
\newcommand{\jec}{\mathcal{J}_{E_c}}
\newcommand{\jix}{\mathcal{J}_{I_X}}
\newcommand{\jex}{\mathcal{J}_{E_X}}
\newcommand{\J}{\mathcal{J}}
\newcommand{\Khc}{\mathcal{K}_{\rm hc}}
\newcommand{\lind}{\mathcal{L}}
\newcommand{\Linv}{\mathcal{L}_{0}^{-1}}
\newcommand{\tr}[1]{\operatorname{tr}\bigl\{#1\bigr\}}
\newcommand{\trb}[1]{\operatorname{tr}\Bigl\{#1\Bigr\}}
\newcommand{\curr}{\mathcal{I}_{\rm ex}}
\newcommand{\iu}{{i\mkern1mu}}
\newcommand{\overbar}[1]{\mkern 1.5mu\overline{\mkern-1.5mu#1\mkern-1.5mu}\mkern 1.5mu}
\newcommand{\ashcomm}[1]{{\color{blue}#1}}

\mathchardef\ordinarycolon\mathcode`\:
\mathcode`\:=\string"8000
\begingroup \catcode`\:=\active
  \gdef:{\mathrel{\mathop\ordinarycolon}}
\endgroup

\begin{document}
\title{Time-resolved Stochastic Dynamics of Quantum Thermal Machines}

\author{Abhaya S. Hegde}
\email{a.hegde@rochester.edu}
\affiliation{
Department of Physics and Astronomy, University of Rochester, Rochester, New York 14627, USA
}

\author{Patrick P. Potts}
\email{patrick.potts@unibas.ch}
\affiliation{Department of Physics and Swiss Nanoscience Institute, University of Basel, Klingelbergstrasse 82 CH-4056, Switzerland}

\author{Gabriel T. Landi}
\email{gabriel.landi@rochester.edu}
\affiliation{
Department of Physics and Astronomy, University of Rochester, Rochester, New York 14627, USA
}

\begin{abstract}
Steady-state quantum thermal machines are typically characterized by a continuous flow of heat between different reservoirs.
However, at the level of discrete stochastic realizations, heat flow is unraveled as a series of abrupt quantum jumps, each representing an exchange of finite quanta with the environment.
In this work, we present a framework that resolves the dynamics of quantum thermal machines into cycles classified as engine-like, cooling-like, or idle.
We analyze the statistics of individual cycle types and their durations, enabling us to determine both the fraction of cycles useful for thermodynamic tasks and the average waiting time between cycles of a given type.
Central to our analysis is the notion of intermittency, which captures the operational consistency of the machine by assessing the frequency and distribution of idle cycles.
Our framework offers a novel approach to characterizing thermal machines, with significant relevance to experiments involving mesoscopic transport through quantum dots.
\end{abstract}

\maketitle

\begin{figure}[t]
    \centering
    \includegraphics[width=\columnwidth]{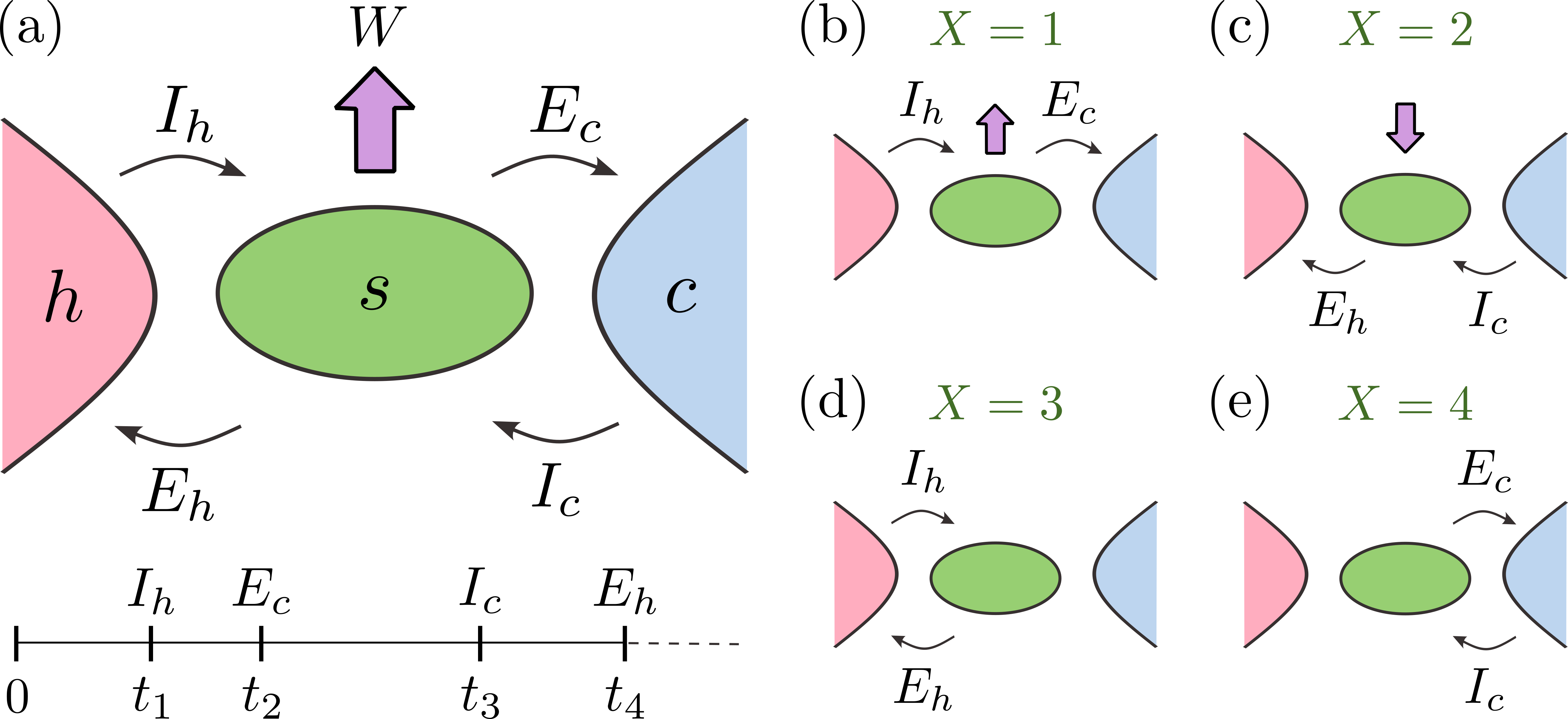}
    \caption{{\bf (a)} In a quantum thermal machine ($s$), heat injection ($I$) and extraction ($E$) are mediated by hot ($h$) and cold ($c$) reservoirs, represented as random events occurring at random times within the quantum jump unraveling. 
    {\bf (b-e)} These jumps can be categorized into four cycles denoted by pairs $I_\bullet E_\bullet$ and labeled by $X$.
    {\bf (b)} Work extraction cycle ($X = 1$): Heat is transferred from hot to cold bath, extracting work.
    {\bf (c)} Cooling cycle ($X = 2$): Excitations move from cold to hot bath, consuming work. These are useful cycles.
    {\bf (d, e)} Idle cycles ($X = 3, 4$): No heat transfer occurs overall.}
    \label{fig:system}
\end{figure}

\textit{Introduction.---}
A typical quantum thermal machine consists of a system situated between hot and cold thermal baths, extracting or absorbing energy in the form of work, as depicted in Fig.~\ref{fig:system}~\cite{Binder18,Kosloff14,Bhattacharjee21,Mitchison19,Sai16}.
As an engine, it extracts work while transferring heat from hot to cold; as a refrigerator, it absorbs work to move heat from cold to hot.
In autonomous machines, this is usually pictured as a continuous process, where heat and work constantly flow through the system~\cite{Niedenzu19, Hammam21, Roulet17}.
However, within the microscopic domain, the stochastic nature of system and bath interactions endows an alternative perspective where energy is exchanged with the baths in the form of abrupt jumps. 
This is the basis for stochastic thermodynamics in classical (Pauli) rate equations~\cite{Pauli28, Seifert19, Harris07}, as well as quantum models in the quantum jumps formalism~\cite{Carmichael93, Dalibard92, Garrahan10, Donvil2022, Breuer03, Molmer93, Pekola13, Manzano22, Leggio13, Wiseman09, Knight98, Landi24}.

The jumps occur at random times and in random ``channels.''
Let us broadly classify these channels as either an injection $(I)$ or an extraction $(E)$ of energy into or out of a system induced by hot $(h)$ or cold $(c)$ baths, resulting in four distinct types of monitored channels $\mathbb{M} = \{I_h, E_h, I_c, E_c\}$. 
Generalizing to multiple injection and extraction channels per bath is straightforward.
The quantum trajectory of such a machine, in the quantum jump unraveling, appears as a random string, e.g.,
\begin{equation}\label{string}
    I_h E_c I_c I_h E_h E_c I_h I_c E_h I_c \ldots,
\end{equation}
along with their timestamps $t_1, t_2, \ldots$, indicating when each jump occurred. 
This representation of the dynamics is grounded in several experimental observations either through a direct detection of jumps~\cite{Hofmann2016, Minev2019, Murch21} or by monitoring the states continuously to deduce the jump processes driving the observed state transitions~\cite{Bergquist86, Bylander05, Fujisawa04, Fujisawa06, Gleyzes07, Gustavsson06, Gustavsson09, Lu03, Ueda16}.
Note that only heat exchange events with the environment are included in~\eqref{string}, as work events, typically associated with unitary drives are assumed undetectable~\cite{Binder18}.

The central question we address in this work is \emph{can specific thermodynamic cycles be identified solely from strings like~\eqref{string} so that their statistics can be explored?}
For instance, one might intuitively characterize the sequence $I_c E_h$ as a refrigeration cycle, since an energy quanta was injected from $c$ to the system, and subsequently extracted to $h$ suggesting work consumption.
Similarly, $I_h E_c$ could be seen as an engine-like process (or accelerator~\cite{dechiara18}).
These are both examples of ``useful cycles.''
Conversely, pairs such as $I_h E_h$ and $I_c E_c$ are events that fail to peddle quanta of energy overall, and incur no entropy production. We refer to these as ``idle cycles".
While a machine might operate as an engine \emph{on average}, the stochastic nature of these processes manifests in individual realizations yielding different cycles~\cite{Andrew21, Monsel25}.

Classifying cycles raises several meaningful questions, such as
\begin{itemize}\itemsep-0.1cm
    \item What is the probability of each type of cycle?
    \item How are cycles related to steady-state currents?
    \item What is the time required to complete each cycle?
    \item How many idle cycles precede a useful one?
\end{itemize}
These questions relate to the extensive literature on Full Counting Statistics (FCS)~\cite{Esposito07,Brandes08,Campisi11}, fluctuation theorems~\cite{Esposito09, Seifert12, Jarzynski04, Jarzynski11, Andrieux09, Campisi14, Campisi15, Miller21}, and thermodynamical aspects of quantum trajectories~\cite{Igor10,Karimi20,Paul20}.
Addressing them involves exploring time-resolved and cycle-resolved quantities, offering a fine-grained understanding of the dynamics.

In attempting to classify cycles this way, a challenge arises when the system can withhold multiple excitations at once. 
For instance, in the string~\eqref{string}, what meaning should be ascribed to the substring $I_c I_h E_h E_c$?
Because excitations are indistinguishable, it is impossible to infer if this was 
$\wick{\c1 I_c \c2 I_h \c2 E_h \c1 E_c}$ (two idles) or
$\wick{\c1 I_c \c2 I_h \c1 E_h \c2 E_c}$ (a refrigeration followed by an engine cycle).
While this is not an issue as far as the average heat and work currents are concerned, it does cause ambiguity in defining time-resolved quantities.
In this letter, we focus on systems that can retain only one excitation at a time; i.e., when injections and extractions alternate ($I_\bullet E_\bullet I_\bullet E_\bullet \ldots$) in the trajectory.
This assumption is common in experiments involving single~\cite{Gustavsson06,Gustavsson09,Lu03} or double~\cite{Wiel02,Thierschmann13,Thierschmann15,Jaliel19,Khan2021} quantum dots in the Coulomb blockade regime, as well as realizations of quantum heat engines~\cite{Blickle12,Zhang22}.

First, we establish the restrictions imposed by the single-excitation hypothesis on a quantum Markovian master equation.
Then, we employ the tools of waiting time distributions of the quantum jump unraveling~\cite{Knight98,Wiseman09,Landi24} to fully characterize the statistics of cycles. 
Finally, we illustrate our results with a three-level maser example. 

\textit{Theory.---}
We consider a finite-dimensional system weakly coupled to hot and cold baths.
Work may be performed either by a driven Hamiltonian $H(t)$ or by additional work reservoirs. 
It is assumed that the dynamics can be described by a quantum master equation~\cite{GKS76, Lindblad76} ($\hbar = k_B = 1$ throughout),
\begin{align}\label{GKSL}
 \frac{d\rho}{dt} = \mathcal{L}_t\rho\equiv& -i[H(t), \rho] + \sum_n  \mathcal{D}[K_n] \rho \\
 \nonumber &+\sum_{\alpha\in\{h,c\}, j}\Bigl(\gamma_{\alpha j}^{-} \mathcal{D}[L_{\alpha j}] + \gamma_{\alpha j}^{+} \mathcal{D}[L_{\alpha j}^\dagger]\Bigr)\rho ,
\end{align}
where $\mathcal{D}[L]\rho = L \rho L^\dagger - \frac{1}{2}\{L^\dagger L, \rho\}$.
Here, $\{L_{\alpha j}\}$ are jump operators for the hot ($\alpha = h$) and cold ($\alpha = c$) baths, with 
$L_{\alpha j}$ denoting extractions and $L_{\alpha j}^\dagger$ denoting injections, each occurring at rates $\gamma_{\alpha j}^{\mp}$, respectively.
Finally, $K_n$ are jump operators of work reservoirs, which are often used in describing absorption refrigerators~\cite{Kosloff12, Juan13, Correa14, Maslennikov19, Manikandan20}.

We assume one can only monitor whether energy is injected (extracted) from (to) the hot or cold baths without identifying the specific jump operator (indexed by $j$) responsible. Therefore, the four corresponding jump superoperators are
\begin{equation}
\label{jumpops}
    \mathcal{J}_{E_\alpha}\rho = \sum_j \gamma_{\alpha j}^- L_{\alpha j}^{} \rho L_{\alpha j}^\dagger,
    \;\;
    \mathcal{J}_{I_\alpha}\rho = \sum_j \gamma_{\alpha j}^+ L_{\alpha j}^\dagger \rho L_{\alpha j}^{}.
\end{equation}

As our first result, we prove in the Supplemental Material~\cite{supp} that the condition for the quantum trajectory to have alternating injections and extractions (i.e. at most a single excitation)  is achieved, if and only if, there exist two subspaces $\he$ and $\hi$ spanning the system Hilbert space $\mathcal{H}$, such that: 
\begin{subequations}
\label{conditions}
\begin{align}
    & L_{\alpha j} = \mathcal{P}_E L_{\alpha j} \mathcal{P}_I \quad \forall \; \alpha,j; \\
    & \mathcal{P}_E H(t) \mathcal{P}_I = \mathcal{P}_I H(t) \mathcal{P}_E =0; \\
    & \mathcal{P}_E K_n \mathcal{P}_I = \mathcal{P}_I K_n \mathcal{P}_E=0,
\end{align}
\end{subequations}
where $\mathcal{P}_{E/I}$ are projection operators onto $\mathcal{H}_{E/I}$, satisfying $\mathcal{P}_E + \mathcal{P}_I = 1$.
Thus, the jump operators of the baths must be block upper-triangular, while those of the work reservoir, and Hamiltonian must be block diagonal in the basis spanned by the states in the subspaces $\mathcal{H}_{E}$ and $\mathcal{H}_I$.

\begin{figure}
    \centering
    \includegraphics[width=\linewidth]{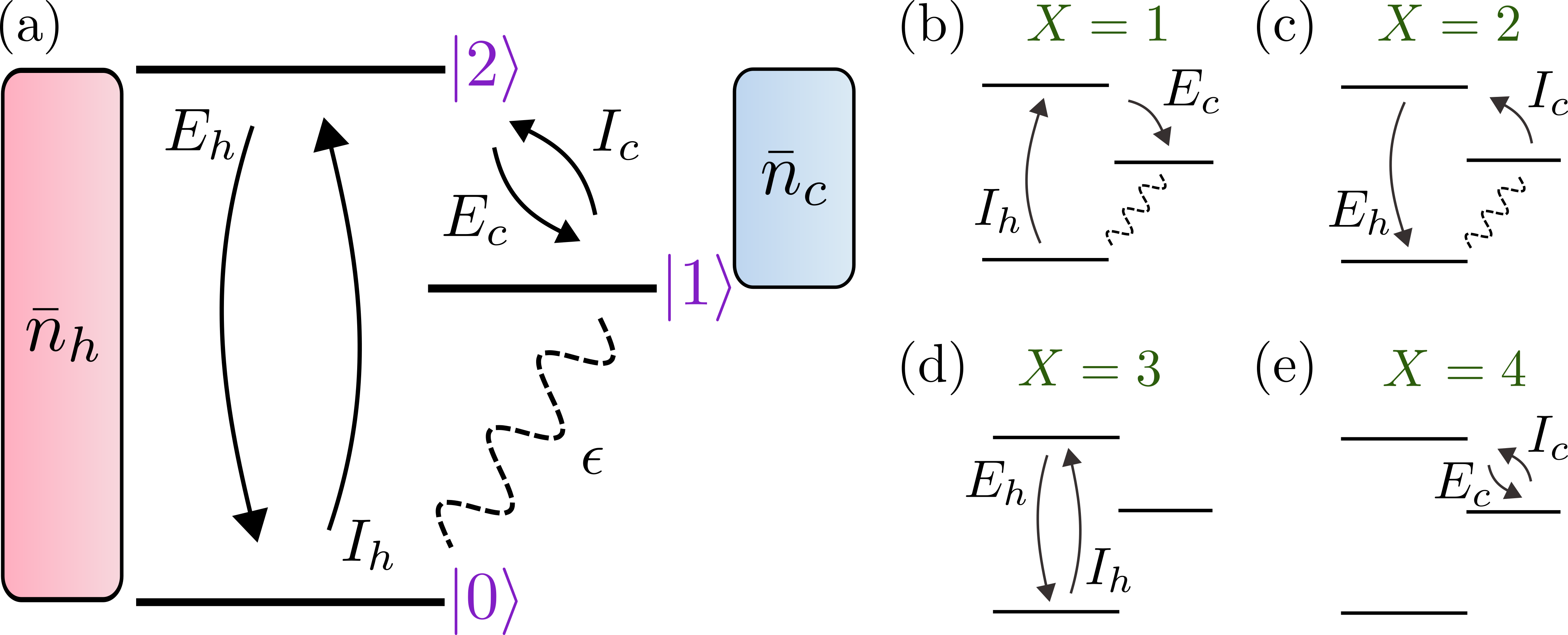}
    \caption{
    {
    \bf (a)} Schematic of a three-level maser connected to hot and cold baths and driven by a Rabi drive, illustrating the four jump processes induced by the baths.
   {\bf (b-e)} All four cycles for this model akin to Fig.~\ref{fig:system}\,(b-e).
    }
    \label{fig:maser_only}
\end{figure}

Consequently, $L_{\alpha j}$ takes the system to $\he$ by extracting energy, while $L_{\alpha j}^\dagger$ directs it to $\hi$ by injecting energy.
We refer to $\he$ and $\hi$ as post-extraction and post-injection subspaces. 
While the unitary dynamics and work reservoirs can inject (extract) work into (out of) the system, this result implies that such processes must occur within each subspace.
Transitions between these subspaces are only feasible through interactions with the baths.
An example is the three-level maser [\ashcomm{see Fig.~\ref{fig:maser_only}\,(a)}]; other examples are hinted in~\cite{supp}.

We henceforth assume, as is often the case, that there exists a rotating frame where $H(t)$ is time-independent, and that the steady-state ($\mathcal{L}\rss=0$) in this frame is unique.
The single-excitation hypothesis implies a conservation law for the average excitation current exchanged with the baths,
\begin{equation}
\label{excitation_current}
    \curr := \tr{ \bigl(\jec - \jic \bigr) \rss } = - \tr{ \bigl(\jeh - \jih \bigr) \rss },
\end{equation}
which is deduced by noting that $\tfrac{d}{dt}\tr{ \mathcal{P}_E \rho(t) \mathcal{P}_E }\to 0$ as the system approaches the steady-state. 
Equation~\eqref{excitation_current} does not imply that the heat currents to both baths are equal, as jumps to each bath generally involve different energies.
Indeed, their mismatch accounts for the work exchanged.
This current is often related to energy fluxes and entropy production rates.

\textit{Statistics of cycles.---}
Under the single-excitation assumption, the trajectories analogous to Eq.~\eqref{string} can be characterized in terms of the statistics of four possible pairs: $I_h E_c$, $I_c E_h$, $I_h E_h$, and $I_c E_c$.
We refer to each pair $I_\bullet E_\bullet$ as a ``cycle'' and label them as $X=1,2,3,4$, respectively (see Fig.~\ref{fig:system}).
$X=1$ is a work extraction cycle~\footnote{Alternatively, it could also represent an accelerator, wherein the direction of heat flow remains the same, but work is injected instead of extracted. For simplicity, we will continue to refer to this as a work cycle. This scenario always holds when the energy exchanged with the cold bath is smaller than that with the hot bath, an assumption we make implicitly.} and $X=2$ a refrigeration cycle, while $X=3, 4$ are idle cycles.

We are interested in the long-time steady-state behavior of strings of the form $X_1X_2\ldots = I_\bullet E_\bullet I_\bullet E_\bullet \ldots$, adopting the convention that strings always begin with an injection. 
Then, as explained in~\cite{supp}, the probability of observing a specific sequence $X_1,\ldots,X_n$, with durations $\tau_1,\ldots,\tau_n$, is given by
\begin{equation}\label{multprob}
p_{X_1,\ldots,X_n}^{}\!(\tau_1,\ldots,\tau_n) = \tr{ \mathcal{O}_{X_n,\tau_n} \ldots \mathcal{O}_{X_1,\tau_1}\pi_E },
\end{equation}
where
\begin{equation}\label{cycleop}
\mathcal{O}_{X,\tau} \equiv \int_0^\tau dt~\jex e^{\mathcal{L}_0(\tau-t)} \jix e^{\mathcal{L}_0t},
\end{equation}
and $\mathcal{L}_0 = \lind - \sum_\alpha (\mathcal{J}_{E_\alpha} + \mathcal{J}_{I_\alpha})$ is the no-jump superoperator. 
In Eq.~\eqref{multprob}, we have introduced the jump steady-state~\cite{supp, Landi23} 
\begin{equation}
\label{jsse}
\pi_E = \frac{(\jeh + \jec)\rss}{\tr{(\jeh + \jec)\rss}} \in \he,  
\end{equation}
to ensure the jump sequence is stationary.

Marginalizing Eq.~\eqref{multprob} over all $(X_i,\tau_i)$ except one yields the probability that a single cycle is of type $X$ and duration $\tau$,
\begin{equation}\label{probxt}
p_X^{}(\tau) = \tr{ \mathcal{O}_{X,\tau}^{}\pi_E }.
\end{equation}
Integrating over $\tau$ yields the  probability that the cycle is of type $X$:
\begin{equation}\label{proball}
p_X^{} = \int_0^\infty d\tau~p_X^{}(\tau) = \tr{\mathcal{O}_X \pi_E},
\end{equation}
where $\mathcal{O}_X = \int_0^\infty d\tau\,\mathcal{O}_{X,\tau} =  \jex \Linv \jix \Linv$,
with $\sum_{X = 1}^{4}p_X^{} = 1$.

The average cycle time given it is of type $X$ reads as
\begin{equation}\label{expect}
E(\tau \mid X) = \frac{1}{p_X^{}}\int_0^\infty d\tau~\tau\,p_X^{}(\tau).
\end{equation}
In~\cite{supp}, we show
\begin{equation}\label{uncond}
E(\tau) = \sum_{X=1}^4 E(\tau \mid X) p_X^{} = \frac{2}{\Khc},
\end{equation}
where $\Khc$ is the dynamical activity of the baths representing the average number of jumps per unit time in the steady-state.
This activity is closely tied to the kinetic uncertainty relation~\cite{Terlizzi18}, and for a classical Markov process, it also relates to information geometry~\cite{Hasegawa24}.

The probabilities in Eq.~\eqref{proball} represent the relative occurrence of each cycle type over many trajectories, regardless of their duration.
In~\cite{supp}, we prove that $p_{1/2}$ and the excitation current from Eq.~\eqref{excitation_current} are related by 
\begin{equation}
\label{prob_curr}
\curr = \frac{p_1 - p_2}{E(\tau)},
\end{equation}
which provides a fundamental connection between usual steady-state currents and our results: 
the system functions as an engine when $p_1 > p_2$, and as a refrigerator when $p_1 < p_2$.

\textit{Example: Three-level system.---}
We apply our results to a three-level maser~\cite{Scovil59,Tannor06,Binder18,Cangemi23,Segal21,Singh20,Kosloff94,Boukobza07,Manzano24,Kalaee2021} whose schematic is depicted in Fig.~\ref{fig:maser_only}. It is coupled to hot and cold baths at energy $\omega_\alpha$ and temperature $T_\alpha$ with their populations following a Bose-Einstein distribution given by $\bar{n}_{\alpha}=[\exp(\omega_{\alpha}/T_\alpha)-1]^{-1}$. 
The maser is driven by the Hamiltonian $H(t) = (\omega_h-\omega_c)\sigma_{11} + \omega_h \sigma_{22} + \epsilon(e^{i\omega_d t} \sigma_{01} + e^{-i\omega_d t}\sigma_{10})$
with a Rabi drive of strength $\epsilon$ and frequency $\omega_d$. 
The jump operators are 
$L_{h} = \sigma_{02}$, $L_{c} = \sigma_{12}$ (and $K_n = 0$) with rates 
$\gamma_{\alpha}^{-} = \gamma_{\alpha}(\bar{n}_\alpha + 1)$ and $\gamma_{\alpha}^{+} = \gamma_{\alpha}\bar{n}_\alpha$.
Here, $\sigma_{ij} = |i\rangle\langle j|$ are the transition operators. The post-extraction subspace is spanned by $\{|0\rangle,|1\rangle\}$, and the post-injection by $\{|2\rangle\}$. As anticipated, the Hamiltonian is block diagonal in the joint basis of these subspaces.

\begin{figure*}
    \centering
    \includegraphics[width=\textwidth]{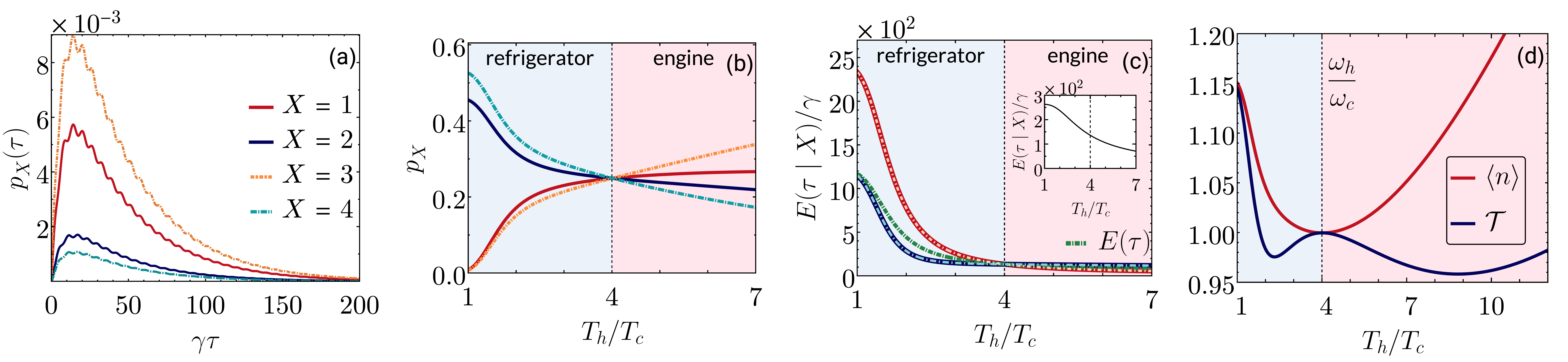}
    \caption{
    {\bf (a-d)} Statistics of cycles in three-level maser from Fig.~\ref{fig:maser_only}.
    {\bf (a)} Probability of observing a cycle $X$ within a duration $\tau$ [Eq.~\eqref{probxt}] at resonance $\omega_d = \omega_h - \omega_c$ and $T_h/T_c = 10$.
    {\bf (b)} Total probability of observing a cycle $X$ [Eq.~\eqref{proball}] and {\bf (c)} expectation values for cycle duration [Eqs.~\eqref{expect}, \eqref{uncond}] as a function of the ratio of bath temperatures.
    A vertical line at $T_h/T_c = \omega_h/\omega_c$ separates the refrigerator and engine regimes. 
    The inset shows all expectation values nearly converge at resonance.
    {\bf (d)} Mean of intervening idle cycles between useful cycles and ratios of fraction of idle-to-useful times against bath gradient.
    The parameters are fixed (in units of $T_c = 1$) at $\gamma_h = \gamma_c \equiv \gamma = 0.05$, $\omega_h = 8$, $\omega_c = 2$, $\omega_d= 4$, $\epsilon = 0.5$ unless mentioned otherwise.
    }
    \label{fig:maser_stats}
\end{figure*}

Figure \ref{fig:maser_stats}\,(a) illustrates $p_X(\tau)$ from Eq.~\eqref{probxt} [see Ref.~\cite{supp} for explicit expressions].
For large $\tau$, these probabilities scale as
\begin{equation}
\label{asymp_pxt}
p_X(\tau) \sim e^{-\Gamma \tau} \Bigl[ 1 + C_X^{} \cos\Bigl( 2 \tau \sqrt{\epsilon^2 + \tfrac{\Delta^2 + \Lambda^2}{4}}+ \phi_X^{} \Bigr) \Bigr],
\end{equation}
where $\Gamma = (\bar{n}_h \gamma_h + \bar{n}_c \gamma_c)/2$ is the net decoherence rate, $\Lambda = (\bar{n}_h \gamma_h - \bar{n}_c \gamma_c)/2$ indicates the bias, $\Delta = (\omega_h-\omega_c) - \omega_d$ is the detuning, and $C_X^{}, \phi_X^{}$ are constants determined by parameters of this model. 
The oscillatory behavior reflects the coherent drive, reminiscent of Rabi oscillations between $|0\rangle$ and $|1\rangle$, while the exponential decay captures the stochastic nature of jump events.  

The marginals $p_X$ from Eq.~\eqref{proball} are shown in Fig.~\ref{fig:maser_stats}\,(b) as a function of the ratio $T_h/T_c$.
The plot highlights the different regimes of operation, which changes from refrigeration to engine at $T_h/T_c = \omega_h/\omega_c$.
It is noteworthy that~\cite{supp}
\begin{equation}
p_1 - p_2 \propto \bar{n}_h - \bar{n}_c, \qquad 
\frac{p_3}{p_1} = \frac{p_2}{p_4} = \frac{(\bar{n}_h + 1) \gamma_h}{(\bar{n}_c + 1) \gamma_c},
\end{equation}
implying that, when $\gamma_h = \gamma_c$, the probabilities of idle cycles bound those of useful ones  across all parameter ranges.
As a result, it is always more likely to observe the machine undergoing a cycle with no net heat transfer.

The average cycle durations [Eqs.~\eqref{expect}, \eqref{uncond}] are plotted in Fig.~\ref{fig:maser_stats}\,(c); for this model, it turns out that $E(\tau \mid 1) = E(\tau \mid 3)$ and $E(\tau \mid 2)=E(\tau \mid 4)$. 
Noticeably, the cycles tend to take much longer in the refrigeration regime.  
Moreover, at resonance, all conditional averages tend to become very close (although not strictly equal), as shown in the inset of Fig.~\ref{fig:maser_stats}\,(c). 
For small $\epsilon$, the probability of useful cycles $p_u := p_1 + p_2$ scales as $\epsilon^2/(\Gamma^2 + \Delta^2)$. 
This highlights that stronger pumps, more resonant drives and lower damping favor useful cycles.

In this model, the excitation current from Eq.~\eqref{prob_curr} is directly related to the steady-state heat, work, and entropy production currents, as
$\dot{Q}_h = \omega_h \curr$, $\dot{Q}_c = - \omega_c \curr$,  $\dot{W} = \omega_d \curr$ and $\dot{\Sigma} = \sigma \curr$, where $\sigma = \omega_c/T_c - \omega_h/T_h$.
The second law $\dot{\Sigma} \geq 0$ confirms the conditions for engine and refrigeration regimes, depending on the sign of $\sigma$.

On the level of individual stochastic events, idle cycles ($X=3,4$) are entropy-neutral, while engine ($X=1$) and refrigeration ($X=2$) cycles produce entropy $\pm \sigma$, respectively. The average entropy produced per cycle is therefore $E(\Sigma_{\rm cyc}) = \sigma (p_1 - p_2)$, which relates to the steady-state entropy production rate as $\dot{\Sigma} = E(\Sigma_{\rm cyc})/E(\tau)$.
The variance in entropy production within each cycle reads
\begin{equation}
{\rm Var}(\Sigma_{\rm cyc}) = \sigma^2 \bigl[(1 - p_{\rm id}) - (p_1 - p_2)^2\bigr],
\end{equation}
where $p_{\rm id} := p_3 + p_4$ is the probability of idle cycles. 
This variance vanishes in the absence of coherent drive ($\epsilon = 0$ implying $p_{\rm id} = 1$, $p_1 = p_2 = 0$) and is bounded by $\sigma^2(1 - p_{\rm id})$ when $p_1 = p_2$.
Thus, the fluctuations in entropy production are directly related to how often the machine fails to produce useful cycles.

\textit{Intermittency of a machine.---}
These findings show that thermodynamic quantities can vary significantly between individual cycles, highlighting the role of the machine's \emph{regularity} or \emph{intermittency} in its performance.
Despite this variability, due to $\dot{Q}_h = \omega_h \curr$ and $\dot{Q}_c = - \omega_c \curr$, these fluctuations leave the steady-state efficiency unaffected, with $\eta = 1 + \dot{Q}_c/\dot{Q}_h = 1 - \omega_c/\omega_h$. 
This perspective aligns with Ref.~\cite{Seifert18}, wherein the need for a complementary metric to characterize small-scale machines was suggested.

Intermittency as a measure should capture the distribution of idle cycles as a proxy for consistency in heat flow.
Concretely, intermittency can be characterized by the average number of idle cycles between two useful ones.
Since the typical thermodynamic variables cannot witness idle cycles, their presence is inferred from only the time the machine spends abstained from transferring heat.
Thus, in a manner analogous to the previous definition, the average fraction of time spent performing idle cycles provides another aspect of intermittency, particularly when idle cycles occur on a different timescale than useful ones.
A perfectly regular machine --- one where only useful cycles occur --- would have zero intermittency.

For the three-level maser, characterizing intermittency is greatly simplified since cycles are independent.
In other words, these cycles form renewal processes.
The trajectory probability from Eq.~\eqref{multprob} factors into a product because the post-injection subspace is a singleton ($|2\rangle$).
The average number of idles $n$ between useful ones, and the fraction representing the average time spent in idle cycles relative to useful cycles $\mathcal{T}$ appear as \cite{supp}
\begin{equation}
\langle n \rangle = \frac{p_{\rm id}}{p_u} = \frac{p_3 + p_4}{p_1 + p_2}, \ \ \
\mathcal{T} = \frac{p_3 E(\tau | 3) + p_4 E(\tau | 4)}{p_1 E(\tau | 1) + p_2 E(\tau | 2)},    
\end{equation}
both of which are plotted in Fig.~\ref{fig:maser_stats}\,(d).
Assuming $\gamma_h = \gamma_c$, we find $\langle n \rangle \geq 1$ and thus the machine operates irregularly.
Selecting an appropriate ratio of bath temperatures, e.g., $T_h/T_c \sim 9$, enables quicker cycle completion but results in a higher participation of idle cycles. This emphasizes the subtle trade-offs involved in balancing two aspects of intermittency.
This trade-off is further illuminated by examining the distributions of idle and useful cycle times as explored in~\cite{supp}.
Moreover, the framework in~\cite{supp} generalizes to machines with correlated cycles, capturing dynamics beyond independent renewal processes.

\textit{Conclusions.---} 
We showed how to unravel the time-dependent statistics of quantum thermal machines, enabling classification of stochastic dynamics into distinct cycles based on how they interact with different resource reservoirs, determination of cycle occurrence frequencies, and cycle durations.
Our results encompass all statistical correlations between cycles, and also connect with known results in FCS for the average excitation current and dynamical activity. 
This approach provides a new avenue for characterizing quantum thermal machines using experimentally accessible data. 
In particular, our formalism could be readily employed to analyze, e.g., mesoscopic transport in quantum dot experiments shedding light on the underlying thermodynamics and emphasizing the role of regularity in heat flow.

A key takeaway from this analysis is the concept of intermittency, i.e.,
the reliability of a machine in performing thermodynamically useful tasks.
Since our approach enables the identification of both useful and idle cycles, we now have the tools to optimize the intermittency for fixed efficiency and output power.
Our results also allow us to examine cycle ``bunching'', specifically how the occurrence of one useful cycle influences the probability of observing another.
These insights have the potential to significantly deepen our understanding and interpretation of quantum stochastic processes.

\begin{acknowledgments}
The authors express thanks to Joseph Smiga, Felix Binder, and Guilherme Fiusa for fruitful discussions.
This research is primarily supported by the U.S. Department of Energy (DOE), Office of Science, Basic Energy Sciences (BES) under Award No. DE-SC0025516. P. P. P. acknowledges funding from the Swiss National Science Foundation (Eccellenza Professorial Fellowship PCEFP2\_194268).
\end{acknowledgments}

\bibliography{ref}

%%%%%%%%%%%%% Supplemental material %%%%%%%%%%%%%%%%%%%%%%%%%%
\clearpage
\onecolumngrid
\begin{center}
\textbf{\large Supplemental Material for ``Time-Resolved Stochastic Dynamics of Quantum Thermal Machines"}\\[0.3cm]

Abhaya. S. Hegde$^{1,*}$, Patrick P. Potts$^{2,\dagger}$, Gabriel T. Landi$^{1,^\ddagger}$\\
\textit{$^1$\small Department of Physics and Astronomy, University of Rochester, Rochester, New York 14627, USA}\\
\textit{$^2$\small Department of Physics and Swiss Nanoscience Institute,
University of Basel, Klingelbergstrasse 82 CH-4056, Switzerland}
\end{center}

%%%%%%%%%% Prefix a "S" to all equations, figures, tables and reset the counter %%%%%%%%%%
\setcounter{equation}{0}
\setcounter{figure}{0}
\setcounter{table}{0}
\setcounter{page}{1}
\setcounter{section}{0}
\setcounter{secnumdepth}{1}
\makeatletter
\renewcommand{\thesection}{S\arabic{section}}
\renewcommand{\theequation}{S\arabic{equation}}
\renewcommand{\thefigure}{S\arabic{figure}}
\renewcommand{\theHfigure}{S\arabic{figure}}
\renewcommand{\bibnumfmt}[1]{[S#1]}
\renewcommand{\citenumfont}[1]{S#1}
%%%%%%%%%% Prefix a "S" to all equations, figures, tables and reset the counter %%%%%%%%%%

Sections S1 to S5 prove various results mentioned in the main text, and the last section S6 pertains to the technical details of the three-level maser example.

\section{\label{sm:proof}Proof of the single-excitation constraints on a master equation}
In this section, we prove the necessary and sufficient conditions given in Eq.~\eqref{conditions} on the master equation from  Eq.~\eqref{GKSL} to achieve quantum trajectories that have alternating injections and extractions ($I_{\bullet}E_{\bullet}I_{\bullet}E_{\bullet}\dots$) of excitations in a system.
Specifically, we comment on the structure of Hamiltonian, jump operators, and work reservoirs.
In what follows, without loss of generality, we shall drop the index referring to the bath $\alpha$ for jump operators $L_{\alpha j}$ as it is not relevant for this proof in particular. We first assume that there are no work reservoirs. We also assume that the Hamiltonian is time independent, even though the proof is unaltered otherwise.

We begin by imposing that any two successive extractions out of the system cannot occur.
For any $k, q$ and for all times $t_1, t_2$, we demand
\begin{equation}
\tr{\mathcal{J}_{E_k} e^{\mathcal{L}_0 t_2} \mathcal{J}_{E_q} e^{\mathcal{L}_0 t_1} \rho} = 0,
\end{equation}
and likewise for two consecutive injections. 
Then, the action of a no-jump operator $\mathcal{L}_0$ is equivalent to an effective non-Hermitian operator, 
\begin{equation}
\label{sm:non-hermitian}
\mathcal{L}_0 \rho = -i \bigl[H_e \rho - \rho H_e^\dagger\bigr],
\end{equation}
where $H_e = H - i\sum_{j}L_j^{\dagger}L_j^{\phantom\dagger}/2$.
Thus, we can express $e^{\mathcal{L}_0t}\rho = e^{-iH_e t}\rho e^{iH_e^\dagger t}$.
Expanding the extraction superoperators, we obtain
\begin{equation}
\gamma_k\gamma_q \tr{L_k e^{-i H_e t_2} L_q e^{-i H_e t_1} \rho e^{i H_e^{\dagger} t_1} L_q^{\dagger} e^{i H_e^{\dagger} t_2} L_k^{\dagger}} = 0,
\end{equation}
for $\gamma_k, \gamma_q > 0$. 
Since we want this to be true for all times $t_1$ and $t_2$, Taylor expanding $e^{-i H_e t} = \sum_{n=0}^{\infty}(-i H_e t)^n/n!$ then allows us to conclude that this will be possible if and only if all powers of $H_e$ satisfy the condition,
\begin{equation}
\label{condition1}
L_k H_e^{n} L_q = 0, \qquad n = 1,2,\ldots
\end{equation}
as well as 
\begin{equation}
\label{condition2}
L_k L_q = 0 \quad \forall \, k, q.
\end{equation}

\subsection*{Constraints on each jump operator}
Setting $k = q$ in Eq.~\eqref{condition2} renders $L_k^2 = 0$.
Let us temporarily ignore the index for the jump operator.
The condition $L^2=0$ means the jump operators must be nilpotent matrices of order two.
From Ref.~\cite{SHorn85} (p.no.~157) one can show that for any such matrix, it is always possible to find a unitary matrix $V$ such that $L = V \Lambda V^\dagger$, where
\begin{equation}
    \Lambda = \sigma_1 \begin{bmatrix} 0 & 1 \\ 0 & 0 \end{bmatrix} \oplus \cdots \oplus \sigma_r \begin{bmatrix} 0 & 1\\ 0 & 0 \end{bmatrix} \oplus 0_{d - 2r},
\end{equation}
with $\sigma_1 \geq \cdots \geq \sigma_r > 0$ being the positive singular values of $L$ and $\oplus$ denoting direct sum. 
Here, $d={\rm dim}(L)$ and $r = {\rm rank}(L)$.
It follows from the rank-nullity theorem~\cite{SHorn85} that $r \leq \left\lfloor\tfrac{d}{2}\right\rfloor$.
Thus, we can choose $d-r$ orthonormal vectors $\bigl\{|\mu_i\rangle\bigr\}_{i=1}^{d-r}$ 
satisfying $\Lambda |\mu_i\rangle = 0$, as well as $r$ vectors 
$\bigl\{|\nu_j\rangle\bigr\}_{j=1}^{r}$ satisfying $\Lambda |\nu_j\rangle = \sigma_j |\mu_i\rangle$. 

We now introduce a rotated basis defined by $|\epsilon_i\rangle = V |\mu_i\rangle$ and $|\iota_j\rangle = V |\nu_j\rangle$, so that the jump operator $L$ acts in the following way:
\begin{align}
L|\epsilon_i\rangle &= 0, \\
L|\iota_j\rangle &= \sum_{i=1}^{r} c_i |\epsilon_i\rangle, \quad c_i \in \mathbb{R}.
\end{align}
From this, we can form orthogonal projectors, $\mathcal{P}_{E} = {\rm span}\{|\epsilon\rangle\}$
and $\mathcal{P}_{I} = {\rm span}\{|\iota\rangle\}$
that obey $\mathcal{P}_{E} + \mathcal{P}_{I} = 1$.
This reformulation allows us to express the operators as $L = \mathcal{P}_{E} L \mathcal{P}_{I}$.
Thus, we conclude that there is a basis of the Hilbert space in which each jump operator is written as a strictly upper-triangular matrix.

To further clarify the procedure, 
let us consider an explicit example for $d = 5$ with $r = 2$ case. According to our proof, such a jump operator $L$ must be unitarily similar to
\begin{equation*}
\Lambda = 
\begin{bmatrix}
0 & \sigma_1 & 0 & 0 & 0 \\
0 & 0 & 0 & 0 & 0 \\
0 & 0 & 0 & \sigma_2 & 0 \\
0 & 0 & 0 & 0 & 0 \\
0 & 0 & 0 & 0 & 0
\end{bmatrix},
\end{equation*}
where $\sigma_1 \geq \sigma_2 > 0$ are positive singular values of $L$. As one may verify, we can find the orthonormal set of vectors and the action of $\Lambda$ on them as,
\begin{align*}
\left.\begin{aligned}
|\mu_1\rangle &= (1, 0, 0, 0, 0)^{T} \\ 
|\mu_2\rangle &= (0, 0, 1, 0, 0)^{T} \\
|\mu_3\rangle &= (0, 0, 0, 0, 1)^{T}
\end{aligned}\right\} \Lambda |\mu_i \rangle &= 0 \text{ for }i = 1, 2, 3,\\
\left.\begin{aligned}
|\nu_1\rangle &= (0, 1, 0, 0, 0)^{T} \\ 
|\nu_2\rangle &= (0, 0, 0, 1, 0)^{T} 
\end{aligned}\right\} \Lambda |\nu_i \rangle &= \sigma_i |\mu_i\rangle \text{ for }i = 1, 2.
\end{align*}
Now, the exact forms of unitary matrices $V$ are dependent on $L$.

\subsection*{Joint injection/extraction subspaces for the entire set of jump operators}

We now return to the full set of jump operators $L_k$.
As just shown, to each one we can associate two orthonormal projectors $\mathcal{P}_{E_k}$ and $\mathcal{P}_{I_k}$ such that $L_k = \mathcal{P}_{E_k} L_k \mathcal{P}_{I_k}$, where each pair satisfies $\mathcal{P}_{E_k} + \mathcal{P}_{I_k} = 1$.
Next, we show that all $\mathcal{P}_{E_k}$ must coincide. 
This follows from the additional condition supplied by Eq.~\eqref{condition2}.

Since for any pair $L_k L_q = (\mathcal{P}_{E_k} L_k \mathcal{P}_{I_k})(\mathcal{P}_{E_q} L_q \mathcal{P}_{I_q}) = 0$, this must imply that $\mathcal{P}_{I_k}\mathcal{P}_{E_q} = (1 - \mathcal{P}_{E_k})\mathcal{P}_{E_q} = 0$.
Thus $\mathcal{P}_{E_q} = \mathcal{P}_{E_k}\mathcal{P}_{E_q}$, allowing us to conclude that $\mathcal{P}_{E_q} = \mathcal{P}_{E_k}$. 
This also implies that $\mathcal{P}_{I_q} = \mathcal{P}_{I_k}$ leading to only two unique projectors  $\mathcal{P}_E$ and $\mathcal{P}_I$ which must be the same for all jump operators associated to the hot and cold baths.

\subsection*{Hamiltonian and work reservoir operators}

From Eq.~\eqref{condition1}, setting $n = 1$  we have that $L_k H L_q = 0$. This implies $\Bigl(\mathcal{P}_E L_k \mathcal{P}_I\Bigr) H \Bigl(\mathcal{P}_E L_q \mathcal{P}_I\Bigr) = 0$ which reduces to $\mathcal{P}_I H \mathcal{P}_E = 0 = \mathcal{P}_E H \mathcal{P}_I$. This completes the proof for the case where $K_n = 0$ and the Hamiltonian is time-independent.

The steps proceed in analogously for time-dependent Hamiltonian too. Since the work reservoir jump operators also do not induce any transitions between the subspaces $\he$ and $\hi$, it follows that $\mathcal{P}_I K_n \mathcal{P}_E = 0 = \mathcal{P}_E K_n \mathcal{P}_I$.

The converse holds true trivially; i.e., given the above forms for the jump operators and the Hamiltonian, it is straightforward that conditions listed in Eq.~\eqref{conditions} are satisfied.

\subsection*{Other schematic examples of thermal machines satisfying our condition}
\begin{figure}[t]
    \centering
    \includegraphics[width=.8\textwidth]{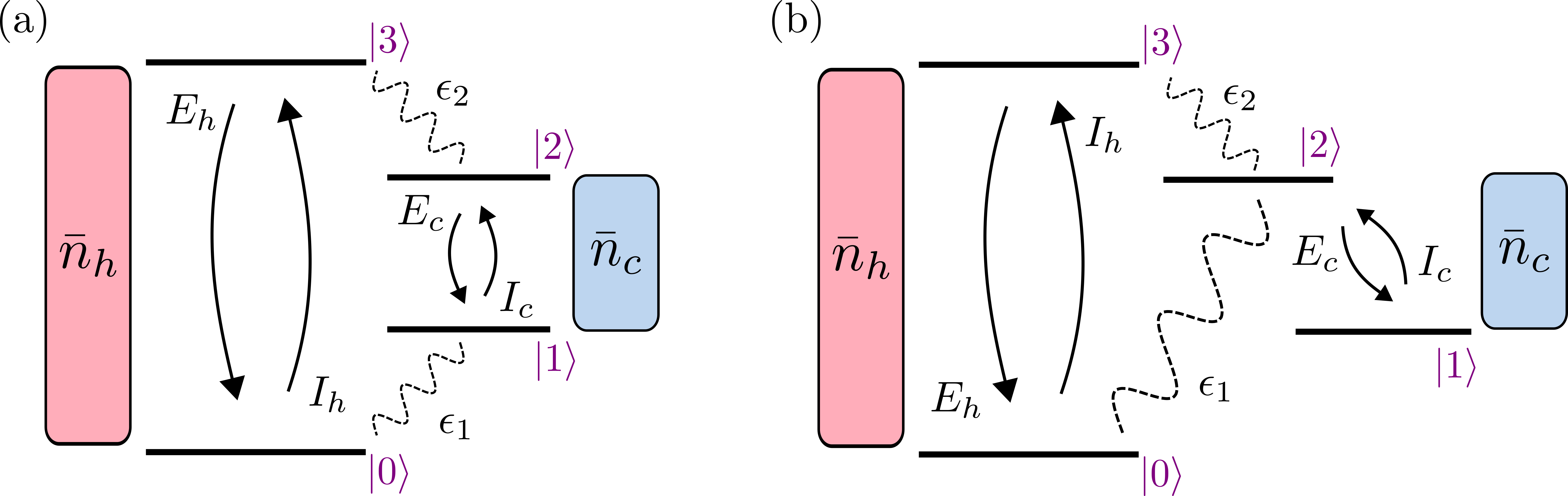}
    \caption{(a) Illustration of a four-level system that satisfies the constraints listed in Eq.~\eqref{conditions} leading to the presence of at most a single excitation at all times. The post-extraction subspace is spanned by the states $|0\rangle$ and $|1\rangle$, while the post-injection subspace consists of states $|2\rangle$ and $|3\rangle$.
    (b) The post-extraction and post-injection subspaces are linked through a Rabi drive which renders it incompatible with our constraints on the master equation and thus may end up having multiple excitations.}
    \label{fig:other_example}
\end{figure}

A key insight to take away is that the Hamiltonian cannot connect post-injection and post-extraction subspaces for the single-excitation assumption to hold.
To clarify the restrictions imposed by the assumption from a physical point of view, we show in Fig.~\ref{fig:other_example} a schematic of a four-level system for which the restriction is applicable and another example for which it is not. 

\section{\label{sm:cycle_stats}Cycle statistics in the long-time limit}
For a given initial state $\rho$, the probability of observing a trajectory of the sort given in Eq.~\eqref{string} takes the form~\cite{SLandi24},
\begin{equation}
\label{trajectory_prob}
\mathcal{P}(t_1, k_1, \ldots, t_n, k_n) = \tr{\mathcal{J}_{k_n}e^{\mathcal{L}_0t_n}\cdots\mathcal{J}_{k_1}e^{\mathcal{L}_0t_1}\rho},
\end{equation}
for all $k_i \in \{I_h, E_h, I_c, E_c\}$ representing the monitored channels and $\tau_i \geq 0$ referring to the time interval between the jumps.

For the case of single-excitations, we shall replace $k_i$'s with alternating injections and extractions.
As in the main text, we adopt the convention that the cycle always begins with an injection. 
For each pair $(t_1,t_2)$, $(t_3,t_4), \ldots$ we change variables from ($t_i,t_{i+1})\to (\tau_i,s_i-\tau_{i})$, and then integrate over $s_i$. 
Bundling each pair $I_\bullet E_\bullet$ together then forms a cycle. 
Carrying out this operation in Eq.~\eqref{trajectory_prob} yields
\begin{equation}\label{multprob2}
p_{X_1,\ldots,X_n}^{}\!(\tau_1,\ldots,\tau_n) = \tr{ \mathcal{O}_{X_n,\tau_n} \ldots \mathcal{O}_{X_1,\tau_1}\rho},
\end{equation}
with the cycle superoperators $\mathcal{O}_{X,\tau}$ as defined in Eq.~\eqref{cycleop} such that $X = 1$ (implying the jump sequence $I_h E_c$) refers to $O_{1, \tau} = \int_{0}^{\tau} dt~\jec e^{\mathcal{L}_0 (\tau - t)} \jih e^{\mathcal{L}_0 t}$ and so on. Equation~\eqref{multprob2} differs from Eq.~\eqref{multprob} of the main text only in the choice of initial state which is left arbitrary here. 
Next, we justify for beginning in a specific initial state $\pi_E$.

We are interested in the distribution~\eqref{multprob2} in the long-time limit of operations of the machine. 
Specifically, we consider the machine to be in operation for an extended period before we begin counting cycles.
We therefore need to choose the initial state $\rho$ which makes Eq.~\eqref{multprob2} stationary.
Interestingly, it is not the steady-state of the master equation that possesses this property, but rather the post-extraction part of jump steady-state (JSS)~\cite{SLandi24}.
The steady-state captures how long the system stays in a given corner of Hilbert space, while the JSS captures how \emph{often} it visits that region, an intuition borrowed from classical stochastic processes~\cite{SKasper2024}.

Given the unique steady-state $\lind \rss = 0$, we define the JSS via 
\begin{equation}
\label{jss}
\pi = \frac{\J \rss}{\Khc},
\end{equation}
where 
\begin{equation}
\label{jump_op}
\J = \sum_{\alpha \in \{h, c\}} (\mathcal{J}_{E_\alpha} + \mathcal{J}_{I_\alpha}),
\end{equation}
is the sum of all jump operators of hot and cold baths, and 
\begin{equation}\label{sup_mat_dynamical_activity}
    \Khc = \tr{\J \rss},
\end{equation}
is the dynamical activity of the hot and cold baths together -- i.e., the average number of jumps per unit time in the steady-state.

Because of the single-excitation hypothesis, the JSS in our case has to be block diagonal in the post-injection and post-extraction subspaces 
\begin{equation}
    \pi = p_E^{} \pi_E^{}  + p_I^{} \pi_I^{}.
\end{equation}
To see this, we can expand Eq.~\eqref{jss} to
\begin{align*}
\pi &= \frac{1}{\Khc}\sum_{\alpha \in \{h, c\}} (\mathcal{J}_{E_\alpha} + \mathcal{J}_{I_\alpha}) \rss = \frac{1}{\Khc} \biggl( \sum_{\alpha, j} \gamma_{\alpha j}^{-} L_{\alpha j}^{} \rss L_{\alpha j}^\dagger \biggr)
+  \frac{1}{\Khc} \biggl( \sum_{\alpha, j} \gamma_{\alpha j}^{+} L_{\alpha j}^\dagger \rss L_{\alpha j}^{} \biggr).
\end{align*}
Now, we define the (properly normalized) post-extracted and post-injected parts of the JSS, along with their weights as follows:
\begin{align}
\pi_E &= \frac{\sum_{\alpha, j} \gamma_{\alpha j}^{-} L_{\alpha j}^{} \rss L_{\alpha j}^\dagger}{\sum_{\alpha, j}\gamma_{\alpha j}^{-} \tr{L_{\alpha j}^\dagger L_{\alpha j}^{} \rss}},
\qquad p_E^{} = \frac{1}{\Khc} \biggl(\sum_{\alpha, j}\gamma_{\alpha j}^{-} \tr{L_{\alpha j}^\dagger L_{\alpha j}^{} \rss} \biggr) \equiv \frac{\mathcal{K}_E}{\Khc}, \\
\pi_I &= \frac{\sum_{\alpha, j} \gamma_{\alpha j}^{+} L_{\alpha j}^{\dagger} \rss L_{\alpha j}^{}}{\sum_{\alpha, j}\gamma_{\alpha j}^{+} \tr{L_{\alpha j}^{} L_{\alpha j}^{\dagger} \rss}},
\qquad p_I^{} = \frac{1}{\Khc} \biggl(\sum_{\alpha, j}\gamma_{\alpha j}^{+} \tr{L_{\alpha j}^{} L_{\alpha j}^{\dagger} \rss} \biggr) \equiv \frac{\mathcal{K}_I}{\Khc},
\end{align}
leading to the desired form.

We can think of $\mathcal{K}_E$ and $\mathcal{K}_I$ as the dynamical activities of the $\he$ and $\hi$ subspaces, representing the average number of jumps going into each of them.
As proved in Sec.~\ref{sm:proof}, in case of single-excitation machines, the states are partitioned into mutually exclusive subspaces.
Consequently, the state has an equal probability of visiting either subspace, leading to
\begin{equation}
\label{pi_split}
p_E^{} = p_I^{} = \frac{1}{2},\qquad \mathcal{K}_E = \mathcal{K}_I = \frac{\Khc}{2},\qquad \pi = \frac{\pi_E + \pi_I}{2}.
\end{equation}

We shall now prove that the distribution~\eqref{multprob2} is stationary when the initial state $\rho$ is the post-extraction JSS $\pi_E$. 
The crucial point is that, due to the alternating nature of injections and extractions, the state immediately preceding an injection must lie within the post-extraction subspace.

\subsection*{Proof of stationarity}

To prove stationarity, it suffices to consider a sequence of two cycles of types $X_1$ and $X_2$, each with durations $\tau_1$ and $\tau_2$, respectively. The corresponding trajectory probability~\eqref{trajectory_prob} for the state $\pi_E$ is 
\begin{equation}
\label{stationarity_proof}
p_{X_1,X_2}^{}(\tau_1, \tau_2) = 
\int\limits_0^{\tau_2} dt_2
\int\limits_0^{\tau_1} dt_1
\trb{ \mathcal{J}_{E_{X_2}} e^{\mathcal{L}_0(\tau_2-t_2)} \mathcal{J}_{I_{X_2}} e^{\mathcal{L}_0 t_2} \mathcal{J}_{E_{X_1}} e^{\mathcal{L}_0(\tau_1-t_1)} \mathcal{J}_{I_{X_1}} e^{\mathcal{L}_0 t_1} \pi_E }.
\end{equation}
Marginalizing over all quantities of the first cycle amounts to the following substitution:
\begin{equation}
\sum_{X_1} \int\limits_0^\infty d\tau \int\limits_0^\tau dt~\mathcal{J}_{E_{X_1}} e^{\mathcal{L}_0(\tau_1-t_1)} \mathcal{J}_{I_{X_1}} e^{\mathcal{L}_0 t_1} = \mathcal{M}_E \mathcal{M}_I,
\end{equation}
where $\mathcal{M}_{k} = -\sum_{\alpha \in \{h, c\}} \mathcal{J}_{k_\alpha} \mathcal{L}_0^{-1}$ for $k \in \{I, E\}$. Combining these two operators gives
\begin{equation}
\mathcal{M} = \mathcal{M}_E + \mathcal{M}_I = - \J \Linv,
\end{equation}
where the final equality follows from Eq.~\eqref{jump_op}. Notably, this operator has a special property that only the jump steady-state satisfies~\cite{SLandi23}: 
\begin{equation}
\mathcal{M}\pi = -\frac{\mathcal{J}\mathcal{L}_0^{-1}\mathcal{J}\rss}{\tr{\mathcal{J}\rss}} = -\frac{\mathcal{J}\mathcal{L}_0^{-1}(\mathcal{L}-\mathcal{L}_0)\rss}{\tr{\mathcal{J}\rss}} = \pi,
\end{equation}
because $\mathcal{L}\rss = 0$.
Since $\mathcal{L}_0$ does not induce any transitions, it follows that $\mathcal{M}_I \pi_I = 0 = \mathcal{M}_E \pi_E$ leading to,
\begin{subequations}
\begin{align}
\mathcal{M}_I \pi_E &= \pi_I,\\
\mathcal{M}_E \pi_I &= \pi_E.
\end{align}
\end{subequations}
Thus, marginalizing Eq.~\eqref{stationarity_proof} over the first cycle leaves us with 
\begin{equation}
    \mathcal{M}_E \mathcal{M}_I \pi_E = \mathcal{M}_E \pi_I = \pi_E,
\end{equation} 
i.e., cycling back to where we started.

As a result, we obtain the following, which subsequently proves the stationarity of Eq.~\eqref{multprob}:
\begin{equation}
    \sum_X \int\limits_0^\infty d\tau~ \mathcal{O}_{X,\tau} \pi_E = \pi_E.
\end{equation}
This also ensures the proper normalization in Eq.~\eqref{proball},
\begin{equation}\label{completeness}
    \sum_X \int\limits_0^\infty d\tau~p_X(\tau) = \sum_X p_X = 1.
\end{equation}

\section{\label{sm:moments}Moments of waiting time duration of a cycle}
The $n$-th moment of the waiting time, given a cycle of type $X$ is a generalization of  Eq.~\eqref{expect}, and reads
\begin{equation}
\mathbb{E}(\tau^n \mid X) = \frac{1}{p_X^{}}\int\limits_0^\infty d\tau~\tau^n p_X^{}(\tau). 
\end{equation}
Substituting for the probability from Eq.~\eqref{proball} and changing the integration variable to $u = \tau - t$ results in the following:
\begin{equation}
\begin{split}
\mathbb{E}(\tau^n \mid X) &= \frac{1}{p_X^{}} \int\limits_0^\infty d\tau~\tau^n \int_0^\tau dt~\trb{ \jex e^{\mathcal{L}_0(\tau - t)} \jix e^{\mathcal{L}_0 t} \pi_E } \\
&= \frac{1}{p_X^{}} \int\limits_0^\infty d\tau~\tau^n \int_0^\infty du~\trb{ \jex e^{\mathcal{L}_0u} \jix e^{\mathcal{L}_0(\tau - u)} \pi_E } \Theta(\tau - u) \\
\label{expstep1}
&= \frac{1}{p_X^{}} \int\limits_0^\infty du~\trb{ \jex e^{\mathcal{L}_0u} \jix \left(\int_0^\infty d\tau~\tau^n e^{\mathcal{L}_0\tau} \Theta(\tau - u)\right) e^{-\mathcal{L}_0u} \pi_E },
\end{split}
\end{equation}
where we introduced Heaviside step function $\Theta(\tau - u)$ in the second equality.
Let us first solve the integral in the parentheses,
\begin{equation}
\begin{split}
I_n &\equiv \int_0^\infty d\tau~\tau^n e^{\mathcal{L}_0\tau}\Theta(\tau - u)\\
&= \tau^n \Bigl(-e^{\mathcal{L}_0\tau}\Linv\Bigr)\Theta(\tau - u)\big\rvert_{\tau = 0}^{\infty} + \Linv\left[n \int_0^{\infty}d\tau~\tau^{n-1}e^{\mathcal{L}_0\tau}\Theta(\tau-u) + \int_0^{\infty}d\tau~\tau^n e^{\mathcal{L}_0 \tau} \delta(\tau - u)\right]\\
&= n \Linv I_{n - 1} + u^n \Linv e^{\mathcal{L}_0 u}\\
\end{split}
\end{equation}
where the first term in the second equality vanishes at both the limits $\tau = 0$ and $\tau = \infty$ (the latter is because of the assumption that there are no dark subspaces). Solving this recursion relation yields,
\begin{equation}
I_n = \Biggl[\sum_{i=0}^{n-1}\frac{n!}{(n-i)!}u^{n-i}\mathcal{L}_0^{-(i+1)} - n! \mathcal{L}_0^{-(n+1)}\Biggr]e^{\mathcal{L}_0u},
\end{equation}
Plugging this back into Eq.~\eqref{expstep1}, we find
\begin{equation}
E(\tau^n \mid X) = \frac{1}{p_X^{}} \Biggl[\sum_{i=0}^{n-1} (-1)^{n-i} \frac{n!}{(n-i)!}\trb{\jex\mathcal{L}_{0}^{-(n-i+1)}\jix\mathcal{L}_0^{-(i+1)}\pi_E} - n! \trb{\jex\mathcal{L}_{0}^{-1}\jix\mathcal{L}_0^{-(n+1)}\pi_E}\Biggr].
\end{equation}
For $n = 1$, we obtain the expectation value for the duration of a given cycle $X$,
\begin{equation}
E(\tau \mid X) = - \frac{1}{p_X^{}} \Bigl(\trb{ \jex \Linv \jix \mathcal{L}_0^{-2} \pi_E } + \trb{ \jex \mathcal{L}_0^{-2} \jix \Linv \pi_E } \Bigr).
\end{equation}
On the other hand, the unconditional waiting time is 
\begin{equation}\label{sup_mat_E_tau}
    E(\tau) = \sum_{X=1}^4 p_X E(\tau \mid X) = - \sum_{X = 1}^4 \left(\trb{ \jex \Linv \jix \mathcal{L}_0^{-2} \pi_E } + \trb{ \jex \mathcal{L}_0^{-2} \jix \Linv \pi_E}\right).
\end{equation}

We can relate $E(\tau)$ in Eq.~\eqref{sup_mat_E_tau} to the dynamical activity defined in Eq.~\eqref{sup_mat_dynamical_activity} through the notion of residence times.
We refer $P_{\rm no}(\tau \mid \rho) = \tr{e^{\mathcal{L}_0 \tau}\rho}$ to denote the survival probability that the system remains in a given state $\rho$ for a time $t$.
Since $-dP_{\rm no}/d\tau$ is then the waiting-time distribution, we can determine the average time the system will remain in state $\rho$ before jumping as 
\begin{equation}
E(\tau \mid \rho) = -\int\limits_0^\infty d\tau~ \tau~ \frac{dP_{\rm no}(\tau \mid \rho)}{d\tau} = -\tr{\mathcal{L}_0^{-1} \rho}.
\end{equation}

Specifically, starting in the jump steady-state yields the well-known result that the dynamical activity is the inverse of the average waiting time in the jump steady-state,
\begin{equation}
\label{state_to_dynact}
E(\tau \mid \pi) = - \frac{\tr{\Linv \J \rss}}{\Khc} = - \frac{\tr{\Linv (\lind - \mathcal{L}_0) \rss}}{\Khc} = \frac{1}{\Khc},
\end{equation}
where we used $\lind\rss =0$ and $\tr{\lind(\cdot)} = 0$.

This relation can be extended to the unconditional expectations as well. To this end, we make use of the following properties of the superoperators $\mathcal{M}_E$ and $\mathcal{M}_I$:
\begin{subequations}
\label{eq:Mtrpres}
\begin{align}
    \trb{\mathcal{M}_E \rho} &=  \trb{\mathcal{P}_I \rho},\\ 
    \trb{\mathcal{M}_I \rho} &=  \trb{\mathcal{P}_E \rho}.
\end{align}
\end{subequations}
Here, we show the first of these equations; the second is analogous. Notice that
\begin{equation}
    \trb{\mathcal{M}_E \rho} = -\trb{\mathcal{J}_E\mathcal{L}_0^{-1}\rho} =  -\trb{\mathcal{J} \mathcal{P}_I\mathcal{L}_0^{-1}\rho} = \trb{(\mathcal{L}-\mathcal{L}_0) \mathcal{P}_I\mathcal{L}_0^{-1}\rho} =  \trb{\mathcal{P}_I \rho},
\end{equation}
since $\mathcal{P}_I$ and $\mathcal{L}_0^{-1}$ commute.
Also, recognize that $\mathcal{P}_I \mathcal{M}_I = \mathcal{M}_I$ and $\mathcal{P}_E \mathcal{M}_E = \mathcal{M}_E$.

Employing the above properties, we are now able to prove Eq.~\eqref{uncond} of the main text:
\begin{equation}
\begin{split}
\label{dyn_act}
E(\tau) &= \sum_X \int\limits_0^\infty d\tau~\tau \int\limits_0^\infty dt~\trb{ \jex e^{\mathcal{L}_0 (\tau-t)} \jix e^{\mathcal{L}_0 t} \pi_E }\\
&= \sum_X \int\limits_0^\infty du \int\limits_0^\infty dt~(u+t)\,\trb{ \jex e^{\mathcal{L}_0u} \jix e^{\mathcal{L}_0 t} \pi_E }
 \\
&= -\trb{ \mathcal{M}_E \mathcal{L}_0^{-1} \mathcal{M}_I \pi_E + \mathcal{M}_E \mathcal{M}_I \mathcal{L}_0^{-1} \pi_E }
 \\
&= E(\tau \mid \pi_I)\tr{ \pi_I } + E(\tau \mid \pi_E)\tr{ \pi_E }
 \\
&= 2E(\tau \mid \pi)  \\
&= \frac{2}{\Khc},
\end{split}
\end{equation}
where the second equality is obtained through a substitution $u = \tau - t$ and the last equality follows from Eq.~\eqref{state_to_dynact}.

\newpage
\section{\label{sm:exc_curr_proof}Connection with the average excitation current}
Equation~\eqref{prob_curr} of the main text provides a fundamental connection between the statistics of cycle and the average excitation current, a much more standard measure to characterize the operation of an engine. 
In order to prove that relation  we shall first recast the probabilities from Eq.~\eqref{proball} as,
\begin{equation}
\begin{split}
p_X^{} &= \tr{\jex \Linv \jix \Linv \pi_E}  \\ 
&= 2 \tr{\jex \Linv \jix \Linv \pi}  \\ 
&= \frac{2}{\Khc}\tr{\jex \Linv \jix \Linv \J \rss},
\end{split}
\end{equation}
where we have used Eq.~\eqref{pi_split} in going to the second equality, and Eq.~\eqref{jss} in the third.
Since $E(\tau) = 2/\Khc$ from Eq.~\eqref{dyn_act}, we have 
\begin{equation}
\begin{split}
\frac{p_X^{}}{E(\tau)} &= \tr{\jex \Linv \jix \Linv \J \rss}\\
&= \tr{\jex \Linv \jix \Linv (\lind - \mathcal{L}_0) \rss}\\
&= -\tr{\jex \Linv \jix \rss},
\end{split}
\end{equation}
where we used $\tr{\lind(\cdot)} = 0$ in the last step. Thus, we can express
\begin{equation}
\begin{split}
\frac{p_1 - p_2}{E(\tau)}
&= \tr{\jeh \Linv \jic \rss} - \tr{\jec \Linv \jih \rss} \\
&= \tr{(\lind - \mathcal{L}_0 - \jec - \jih - \jic) \Linv \jic \rss} - \tr{\jec \Linv (\lind - \mathcal{L}_0 - \jec - \jic - \jeh) \rss}\\
&= \tr{\jec \rss} - \tr{\jic \rss} 
\\
&\equiv \curr,
\end{split}
\end{equation}
which proves Eq.~\eqref{prob_curr}. In simplifying the above, we repeatedly used $\tr{\lind(\cdot)} = 0$ and $\mathcal{L} = \mathcal{L}_0 + \jih + \jeh + \jic + \jec$. Notice that the terms implying two consecutive injections or extractions are naturally traceless in our framework.

\section{\label{sm:puse_proofs}Probabilities for observing intermediate cycles between two of the same cycles}
 
Equations~\eqref{probxt} to~\eqref{prob_curr} in the main text describe the statistics of individual cycles.
To quantify the consistency in heat flow of a stochastic machine, we introduced two measures of intermittency: the average number of idle cycles between useful cycles, and the ratio of average idle cycle time to average useful cycle time.
While the latter measure is based solely on individual cycles, further insights can be gained by extending the analysis to examine the occurrence of multiple cycles in specific orders. This is especially insightful when cycles are correlated.

With the full trajectory distribution~\eqref{multprob}, we can expand the analysis to explore how correlations between cycles influence system behavior.
Specifically, we aim to investigate the time-scale disparity between idle and useful cycles by analyzing the average time between two useful cycles, which requires the full distribution of multiple cycle occurrences.
In this section, we construct such distributions in full generality, including cases where cycles are correlated.
In particular, we establish the following two results: 

Recall that the useful cycles are $X=1,2$ and the idle cycles are $X=3,4$. 
Starting with the cycle superoperators in Eq.~\eqref{cycleop}, we define
\begin{equation}
    \mathcal{O}_{\rm u} = \mathcal{O}_1 + \mathcal{O}_2, 
    \qquad 
    \mathcal{O}_{\rm id} = \mathcal{O}_3 + \mathcal{O}_4,
\end{equation}
and similarly for $\mathcal{O}_{{\rm u},\tau}$ and $\mathcal{O}_{{\rm id},\tau}$.
\begin{enumerate}
    \item The probability that there are $n$ idle cycles between two useful ones is given by
    \begin{equation}\label{pusen}
    \mathbb{P}_{\text{u}}(n) = \frac{\tr{ \mathcal{O}_{\rm u} \mathcal{O}_{\rm id}^{n} \mathcal{O}_{\rm u} \pi_E }}{\tr{ \mathcal{O}_{\rm u} \pi_E }}.
    \qquad n = 0,1,2,\ldots
\end{equation}
    \item The probability that it takes a time $t$ between the conclusion of two useful cycles is given by
    \begin{equation}\label{puset}
    \mathbb{P}_{\text{u}}(t) = \int\limits_{-\infty}^\infty \frac{dz}{2\pi}
    \frac{ \tr{ \widetilde{\mathcal{O}}_{{\rm u},z} (1-\widetilde{\mathcal{O}}_{{\rm id},z})^{\!-1} \mathcal{O}_{\rm u}\pi_E } }{ \tr{ \mathcal{O}_{{\rm u}}\pi_E } }e^{\iu z t},
\end{equation}
where $\widetilde{\mathcal{O}}_{X,z} = \int_0^\infty d\tau~e^{-\iu z \tau} \mathcal{O}_{X,\tau}$.
\end{enumerate}

The proof of Eq.~\eqref{pusen} is straightforward.
We are seeking the probability of observing $n$ idle cycles until a next useful one given that we have already observed a useful cycle. Post the first useful cycle, the state is updated to $\mathcal{O}_{\rm u}\pi_E/\tr{\mathcal{O}_{\rm u}\pi_E}$. We then readily obtain,
\begin{equation}
\mathbb{P}_{\rm{u}}(n) = \frac{\tr{\mathcal{O}_{\rm u}\mathcal{O}_{\rm id}^n\mathcal{O}_{\rm u}\pi_E}}{\tr{\mathcal{O}_{\rm u}\pi_E}}.
\end{equation}

To prove Eq.~\eqref{puset} we proceed as follows. 
Suppose we are interested in observing a specific cycle $X = x$ by time $t$ such that we implicitly assume only other cycles $\overbar{X} \ne x$ occur until then.
This cumulative probability $\mathbb{P}_{X = x, t}$ is then an infinite sum where each term defaults to observing $X$ in the last trial while assigning the probability for observing $\overbar{X}$ until the last but one, 
{\allowdisplaybreaks
\begin{align}
\nonumber \mathbb{P}_{X = x}(t) =\,&p_{X = x}(t) \\
\nonumber &+ \sum_{\overbar{X}_1} \int\limits_{0}^{t} dt_1~p_{\overbar{X}_1, X_2 = x}(t_1, t-t_1) \\
\nonumber &+ \sum_{\overbar{X}_1, \overbar{X}_2} \int\limits_{0}^{t}dt_2\int\limits_{0}^{t_2}dt_1~p_{\overbar{X}_1, \overbar{X}_2, X_3 = x}(t_1, t_2 - t_1, t - t_2) \\
&+ \cdots,
\end{align}}
where we have made use of $p_X(t)$ from Eq.~\eqref{probxt}. 
Written in terms of the cycle operators, this becomes
\begin{equation}
\begin{split}
\mathbb{P}_{X = x}(t) =\,&\tr{ \mathcal{O}_{X = x, t} \pi_E } \\
&+\sum_{\overbar{X}_1}\int\limits_0^t dt_1 \trb{ \mathcal{O}_{X_2 = x, t - t_1}\mathcal{O}_{\overbar{X}_1, t_1} \pi_E } \\
&+\sum_{\overbar{X}_1, \overbar{X}_2}\int\limits_0^t dt_2\int\limits_0^{t_2} dt_1 \trb{\mathcal{O}_{X_3 = x, t-t_2} \mathcal{O}_{\overbar{X}_2, t_2 - t_1}\mathcal{O}_{\overbar{X}_1, t_1} \pi_E } \\
&+\cdots.
\end{split}
\end{equation}
We can now recast it in terms of Eq.~\eqref{multprob} after replacing for the duration of cycles, $\tau_i = t_{i} - t_{i-1}$, for $i = 1, 2, \ldots$ with $\tau_1 = t_1$ and $\tau_n = t - t_{n-1}$,
\begin{align}
\mathbb{P}_{X = x}(t) &= \sum_{n=0}^{\infty} \sum_{\overbar{X}_{1}, \cdots, \overbar{X}_{n-1}} \int\limits_0^\infty d\tau_n \cdots d\tau_1 \trb{ \mathcal{O}_{X_n = x, \tau_n}^{} \mathcal{O}_{\overbar{X}_{n-1}, \tau_{n-1}} \cdots \mathcal{O}_{\overbar{X}_1, \tau_1} \pi_E }~\delta\biggl(t - \sum\limits_{n=1}^\infty\tau_i\biggr).
\end{align}
Here we shall make use of the Fourier expansion of delta function, $\delta(t) = \int_{-\infty}^\infty dz\,e^{izt}/2\pi$, and also introduce the Fourier transformed version of the cycle operator,
\begin{equation}
\widetilde{\mathcal{O}}_{X,z} = \int\limits_{0}^{\infty} dt~\mathcal{O}_{X, t}e^{-izt},
\end{equation}
so that we can simplify the above to,
\begin{align*}
\mathbb{P}_{X = x}(t)
&= \sum_{n=0}^{\infty}~\sum_{\overbar{X}_{1}, \ldots, \overbar{X}_{n-1}}~\int\limits_{-\infty}^\infty \frac{dz}{2\pi}\trb{ \widetilde{\mathcal{O}}_{X_n = x, z}^{} \widetilde{\mathcal{O}}_{\overbar{X}_{n - 1}, z}\cdots\widetilde{\mathcal{O}}_{\overbar{X}_{1}, z}\pi_E } e^{izt}.
\end{align*}
Since all \{$\overbar{X}_k$\} for $k = 1, \cdots, n-1$ are identical so that $\overbar{X}_k \equiv \overbar{X}$, we can further reduce the above equation to,
\begin{equation}
\begin{split}
\label{puseful_time_proof}
\mathbb{P}_{X = x}(t)&= \sum_{\{\overbar{X}\}}\,\int\limits_{-\infty}^\infty\frac{dz}{2\pi}\trb{ \widetilde{\mathcal{O}}_{X_n = x, z} \biggl(\sum_{n = 0}^{\infty}\widetilde{\mathcal{O}}_{\overbar{X}, z}^{n - 1}\biggr)\pi_E }e^{\iu zt}, \\
&= \sum_{\{\overbar{X}\}}\,\int\limits_{-\infty}^\infty \frac{dz}{2\pi}\trb{ \widetilde{\mathcal{O}}_{X_n = x, z}\bigl(1 - \widetilde{\mathcal{O}}_{\overbar{X}, z}\bigr)^{\!-1} \pi_E }e^{\iu zt},
\end{split}
\end{equation}
where we used the summation of infinite geometric series: $\sum_{n=0}^\infty x^n = 1/(1-x)$ for $x < 1$.

Given that we begin with a useful cycle, we can update the initial state to $\pi_E \to \mathcal{O}_{\rm u} \pi_E/\trb{ \mathcal{O}_{\rm u} \pi_E }$.
Summing over the idle cycles for $\{\overbar{X}\} = \{3, 4\}$ and over useful cycles for $X = \{1, 2\}$ in Eq.~\eqref{puseful_time_proof}, we obtain the result from Eq.~\eqref{puset},
\begin{equation}
    \mathbb{P}_{\rm{u}}(t) = \frac{1}{\tr{ \mathcal{O}_{{\rm u}} \pi_E }}~\int\limits_{-\infty}^\infty \frac{dz}{2\pi}\,\trb{ \widetilde{\mathcal{O}}_{{\rm u}, z}\bigl(1 - \widetilde{\mathcal{O}}_{{\rm id}, z}\bigr)^{\!-1} \mathcal{O}_{\rm u}\pi_E }e^{izt}.
\end{equation}
The quantities from Eqs.~\eqref{pusen}~and~\eqref{puset} are plotted for the three-level maser in Fig.~\ref{fig:pu}.

Both Eqs.~\eqref{pusen}-\eqref{puset} are normalized:
\begin{equation}
    \sum_n \mathbb{P}_{\rm u}(n) = 1, \qquad \int_0^{t} \mathbb{P}_{\rm u}(t) = 1.
\end{equation}

It is straightforward to build the average number of idle cycles between the useful ones as
$\langle n \rangle = \sum_n n\,\mathbb{P}_{\rm u}(n)$.
The average waiting times between two useful cycles $\gamma \langle t \rangle = \gamma \int_0^\infty dt~t\,\mathbb{P}_{\rm u}(t)$ can elucidate the role of time-scale in the operation of stochastic machines.
These measures are plotted for the three-level maser in Fig.~\ref{fig:pu} of Sec.~S6.

\section{\label{sm:maser_expressions}Expressions for statistical quantities in three-level maser}

In this section, we provide the equations specific to three-level masers for the statistical measures introduced in the main text. We begin with evaluating the essential operators involved in obtaining those measures.
The computational basis is spanned by $|0\rangle, |1\rangle, |2\rangle$.
We use the notation $\sigma_{ij} = |i\rangle\langle j|$ throughout.

\subsection{Time-independent Hamiltonian}
The maser Hamiltonian is
\begin{equation}
H(t) = (\omega_h-\omega_c)\sigma_{11} + \omega_h \sigma_{22} + \epsilon(e^{i\omega_d t} \sigma_{01} + e^{-i\omega_d t}\sigma_{10}).
\end{equation}
Choosing an appropriate rotating frame of reference $X = \omega_d \sigma_{11} + \omega_h \sigma_{22}$ and rotate according to $H_{\rm rot} = U H U^\dagger$ where $U = e^{\iu X t}$ as in Ref.~\cite{SKalaee2021}, we obtain a time-independent Hamiltonian
\begin{equation}
\widetilde{H} = H_{\rm rot} - X = \Delta \sigma_{11} + \epsilon(\sigma_{01} + \sigma_{10}),
\end{equation}
where $\Delta = (\omega_h - \omega_c) - \omega_d$ is the detuning parameter. We use this Hamiltonian for all further evaluations.

\subsection{Jump operators and steady-state}
According to Eq.~\eqref{jumpops} the jump operators for the three-level maser appear as,
\begin{equation}
\label{sm:jumpops}
\begin{alignedat}{2}
\jeh(\rho) &= \gamma_h(\bar{n}_h+1)\sigma_{02}\rho\sigma_{02}^\dagger, \qquad
&&\jih(\rho) = \gamma_h \bar{n}_h \sigma_{20}\rho\sigma_{20}^\dagger,\\[0.2cm]
\jec(\rho) &= \gamma_c(\bar{n}_c+1) \sigma_{12}\rho\sigma_{12}^\dagger, \qquad
&&\jic(\rho) = \gamma_c \bar{n}_c\sigma_{21}\rho\sigma_{21}^\dagger.
\end{alignedat}
\end{equation}
In this case, the jump steady-state along with its post-extracted and post-injected parts are,
\begin{subequations}
\begin{align}
\pi_E &= \biggl( \dfrac{\gamma_c(\bar{n}_c + 1)}{\gamma_h(\bar{n}_h + 1)} + 1 \biggr)^{\!-1} |0\rangle\langle 0| + \biggl( \dfrac{\gamma_h(\bar{n}_h + 1)}{\gamma_c(\bar{n}_c + 1)} + 1 \biggr)^{\!-1} |1\rangle \langle 1|, \\[0.2cm]
\pi_I &= |2\rangle \langle 2|,\\[0.2cm]
\pi &= \frac{\pi_E + \pi_I}{2}.
\end{align}
\end{subequations}
Because there are no work reservoirs, the action of the no-jump superoperator can be described solely in terms of the non-Hermitian Hamiltonian, $H_e = \widetilde{H} - \iu/2 \sum_j L_j^\dagger L_j^{}$. 
For the maser, this reads
\begin{equation}
\label{non-hermitian_ham}
    H_e = \frac{1}{2\iu}\begin{pmatrix}
                        \bar{n}_h \gamma_h & 2\iu\epsilon & 0 \\
                        2\iu\epsilon & \bar{n}_c \gamma_c + 2\iu\Delta & 0 \\
                        0 & 0 & \gamma_c (\bar{n}_c + 1) + \gamma_h (\bar{n}_h + 1)
                        \end{pmatrix}.
\end{equation}
In what follows, we shall make use of the following simplifying variables wherever required:
\begin{subequations}
\label{simple_1}
\begin{align}
    \gamma &= \tfrac{1}{2}\left(\gamma_h + \gamma_c\right),
    \\[0.2cm]
    \Gamma &= \tfrac{1}{2}\left(\bar{n}_h \gamma_h + \bar{n}_c \gamma_c\right),
    \\[0.2cm]
    \Lambda &= \tfrac{1}{2}\left(\bar{n}_h \gamma_h - \bar{n}_c \gamma_c\right).
\end{align}
\end{subequations}

\subsection{Time-resolved probabilities for completing a cycle}
The following denote the time-resolved probabilities $p_X^{}(\tau)$ [see Eq.~\eqref{probxt}] of completing a cycle $X \in \{1, 2, 3, 4\}$ at time $\tau$.
For $X = 1$, substituting the required operators from Eq.~\eqref{sm:jumpops} and Eq.~\eqref{sm:non-hermitian} yields
\begin{equation}
\begin{split}
p_1^{}(\tau) &= \int\limits_0^\tau dt~\trb{ \mathcal{J}_{E_c} e^{\mathcal{L}_0 (\tau-t) }\mathcal{J}_{I_h} e^{\mathcal{L}_0 t} \pi_E } \\
&= \frac{\gamma_h^2 \bar{n}_h(\bar{n}_h + 1) \gamma_c (\bar{n}_c + 1)}{\gamma_h(\bar{n}_h + 1) + \gamma_c (\bar{n}_c + 1)} %
\int\limits_0^\tau dt~ \left\lvert\langle 2| e^{-iH_e(\tau - t)}|2\rangle\langle 0| e^{-i H_e t}|0\rangle\right\rvert^2 \\
&\qquad+ 
\frac{\gamma_h \bar{n}_h \gamma_c^2 (\bar{n}_c + 1)^2}{\gamma_h(\bar{n}_h + 1) + \gamma_c (\bar{n}_c + 1)}%
\int\limits_0^\tau dt~ \left\lvert\langle 2| e^{-iH_e(\tau - t)}|2\rangle\langle 0| e^{-i H_e t}|1\rangle\right\rvert^2.
\end{split}
\end{equation}
A general expression for these probabilities can be written down as a sum of integrals,
\begin{equation}
\label{sm:pxt}
\begin{split}
p_X^{}(\tau) 
&= \frac{A_X}{2(\gamma + \Gamma)} \int\limits_0^\tau dt~ \left\lvert\langle 2| e^{-iH_e(\tau - t)}|2\rangle\langle s_X^{}| e^{-i H_e t}|s_X^{}\rangle\right\rvert^2 \\
&\qquad+
\frac{B_X}{2(\gamma+\Gamma)}\int\limits_0^\tau dt~ \left\lvert\langle 2| e^{-iH_e(\tau - t)}|2\rangle\langle 1|e^{-i H_e t}|0\rangle\right\rvert^2,
\end{split}
\end{equation}
with $s_X^{} = [(-1)^X+1]/2$, which takes the value zero (one) for odd (even) $X$. The constants $A_X, B_X$ for each cycle are defined in Table~\ref{tab:constants}.

\begin{table}[]
\begin{tabular}{@{}llll@{}}
\toprule
\multicolumn{2}{c}{Cycle type} & \multicolumn{1}{c}{$A_X^{}$}                                & \multicolumn{1}{c}{$B_X^{}$}                            \\ \midrule
$X = 1$     & engine           & $\gamma_h^2\gamma_c\bar{n}_h(\bar{n}_h + 1)(\bar{n}_c + 1)$ & $\gamma_h\gamma_c^2\bar{n}_h(\bar{n}_c + 1)^2$          \\ \midrule
$X = 2$     & refrigeration    & $\gamma_h\gamma_c^2(\bar{n}_h + 1)\bar{n}_c(\bar{n}_c + 1)$ & $\gamma_h^2\gamma_c(\bar{n}_h + 1)^2\bar{n}_c$          \\ \midrule
$X = 3$     & hot idle         & $\gamma_h^3\bar{n}_h(\bar{n}_h+1)^2$                        & $\gamma_h^2\gamma_c\bar{n}_h(\bar{n}_h+1)(\bar{n}_c+1)$ \\ \midrule
$X = 4$     & cold idle        & $\gamma_c^3\bar{n}_c(\bar{n}_c+1)^2$                        & $\gamma_c^2\gamma_h\bar{n}_c(\bar{n}_c+1)(\bar{n}_h+1)$ \\ \bottomrule
\end{tabular}
\caption{The constants that enter in Eq.~\eqref{sm:pxt} to obtain time-resolved probabilities.}
\label{tab:constants}
\end{table}
With the help of variables introduced in Eq.~\eqref{simple_1}, it is straightforward to see that
\begin{equation}
\langle 2| e^{-iH_e(\tau - t)}|2\rangle = e^{-(\Gamma+\gamma)(\tau-t)}.
\end{equation}
Now, we seek the exponential of the non-diagonalized $2 \times 2$ block of $H_e$ spanned by states $\{|0\rangle, |1\rangle\}$.
Denoting this block as $H_e'$, it can be expressed as:
\begin{equation}
H_e' = b_0 \mathbb{I} + b_x \sigma_x + b_z \sigma_z,
\end{equation}
for $\mathbb{I} = \begin{pmatrix}1 & 0\\ 0 & 1\end{pmatrix}$, $\sigma_x = \begin{pmatrix} 0 & 1 \\ 1 & 0 \\ \end{pmatrix}$, and $\sigma_z = \begin{pmatrix} 1 & 0 \\ 0 & -1 \end{pmatrix}$ with coefficients:
\begin{subequations}
\begin{align}
b_0 &= \frac{\Delta}{2} - \frac{i}{4} \bigl(\bar{n}_h \gamma_h + \bar{n}_c \gamma_c\bigr) = \frac{1}{2} \bigl(\Delta - i\Gamma\bigr),\\
b_x &= \epsilon,\\
b_z &= -\frac{\Delta}{2} - \frac{i}{4} \bigl(\bar{n}_h \gamma_h - \bar{n}_c \gamma_c\bigr) = -\frac{1}{2}\bigl(\Delta + i \Lambda \bigr).
\end{align}
\end{subequations}
Making use of the properties of Pauli matrices allows for a simpler exponentiation,
\begin{equation}
    e^{-i H_e' t} = e^{-i b_0 t}\biggl[ \cos\bigl(\lvert b \rvert t\bigr) \mathbb{I} - i \sin\bigl(\lvert b \rvert t\bigr) \frac{b_x \sigma_x + b_z \sigma_z}{\lvert b \rvert} \biggr],
\end{equation}
with $b \equiv \lvert b \rvert = \sqrt{b_x^2 + \lvert b_z \rvert^2} = \sqrt{\epsilon^2 + \frac{1}{4}\bigl(\Delta^2 + \Lambda^2 \bigr)}$. Upon substitution, a tedious yet straightforward calculation yields
\begin{equation}
\label{sm:asymp_pxt}
p_X^{}(\tau) = C_1 e^{-\Gamma \tau} + C_2 e^{-\Gamma \tau} \cos(2b\tau + \phi) - C_3 e^{-2(\Gamma + \gamma) \tau},
\end{equation}
wherein these constants are defined for each cycle as follows:
\begin{subequations}
\begin{alignat}{2}    
C_1 &\equiv C_1(a, b, c, d, A_X^{}, B_X^{}) &&= \frac{A_X^{} b^2 + \left(B_X^{} - A_X^{}\right)c^2}{4 a b^2 \left(a + \frac{\Gamma}{2}\right)},\\[0.2cm]
C_2 &\equiv C_2(a, b, c, d, A_X^{}, B_X^{}) &&= \frac{\left\lvert \left(b^2 - a^2\right) \left(A_X^2 b^2 d^2 - \left(A_X^{} - B_X^{}\right)^2 c^4\right) \right\rvert^{\frac{1}{2}}}{4 b^2 \left(a^2 + b^2\right) \left(a + \frac{\Gamma}{2}\right)},\\[0.2cm]
C_3 &\equiv C_3(a, b, c, d, A_X^{}, B_X^{}) &&= -\frac{A_X^{} \left(a^2 + b^2 + a d\right) + \left(B_X^{} - A_X^{}\right)c^2}{4 a \left(a^2 + b^2\right) \left(a + \frac{\Gamma}{2}\right)},\\[0.2cm]
\phi &\equiv \phi(a, b, c, d, A_X^{}, B_X^{}) &&= \tan^{-1}\biggl[\frac{\left(B_X^{} - A_X^{}\right)bc^2 + A_X^{}abd}{2\left(A_X^{}-B_X^{}\right)a^2c^2 + A_X^{}b^2d}\biggr],
\end{alignat}
\end{subequations}
with
\begin{equation}
a = \left(\gamma + \frac{\Gamma}{2}\right),\quad
c = \frac{\epsilon}{\sqrt{2}},\quad
d = \frac{\Lambda}{2}.
\end{equation}

These expressions are plotted in Fig.~\ref{fig:maser_stats}\,(a) of the main text assuming certain parameter values.
In the long time limit $\tau \to \infty$, the last term in Eq.~\eqref{sm:asymp_pxt} vanishes quicker than the other two. Thus, asymptotically the equations resemble the form 
\begin{equation}
p_X \sim C_1 e^{-\Gamma \tau} \left(1 + \frac{C_2}{C_1} \cos(2 b\tau + \phi) \right).
\end{equation}
Identifying $C_X = C_2/C_1$ recovers the scaling mentioned in Eq.~\eqref{asymp_pxt} of the main text.

\subsection{Probability of observing a cycle of each kind irrespective of waiting times}

Marginalizing the previous results over $\tau$ yield the probabilities $p_X$ of observing a cycle of type $X$ [Eq.~\eqref{proball}]. 
In addition to the ones mentioned in Eq.~\eqref{simple_1}, let us simplify the expressions using the following variables:
\begin{subequations}
\begin{align}
    \bar{n}_{\rm ave} &= \dfrac{\Gamma}{\gamma} = \dfrac{\bar{n}_h \gamma_h + \bar{n}_c \gamma_c}{\gamma_h + \gamma_c},
    \\[0.2cm]
    \kappa &= \dfrac{\Gamma^2 - \Lambda^2}{2\Gamma} = \dfrac{\bar{n}_h \bar{n}_c \gamma_h \gamma_c}{\bar{n}_h \gamma_h + \bar{n}_c \gamma_c},
    \\[0.2cm]
    \gamma_{\rm cl} &= \dfrac{2\epsilon^2 \Gamma}{\Delta^2 + \Gamma^2}.
\end{align}
\end{subequations}
The last quantity is the classical transition rate between $|0\rangle$ and $|1\rangle$ derived by assuming Lorentzian broadening in Fermi's golden rule, as explained in Ref.~\cite{SKalaee2021}.
With these definitions, we find
\begin{align}
p_1^{} &= \tr{ \mathcal{J}_{E_c} \mathcal{L}_0^{-1} \mathcal{J}_{I_h} \mathcal{L}_0^{-1} \pi_E }, \nonumber \\[0.2cm]
&=
\frac{\kappa}{2\Gamma}
\frac{\left(1 + \frac{1}{\bar{n}_c}\right)}{\left({1+\frac{1}{\bar{n}_{\rm ave}}}\right)}
\left[\biggl(1 + \frac{\kappa}{\gamma_{\rm cl}}\biggl)^{\!-1}
+
\frac{\left(1 + \frac{1}{\bar{n}_h}\right)}{\left({1+\frac{1}{\bar{n}_{\rm ave}}}\right)}
\biggl(1 + \frac{\gamma_{\rm cl}}{\kappa}\biggl)^{\!-1}\right],
\\[0.3cm]
p_2^{} &= \tr{ \mathcal{J}_{E_h} \mathcal{L}_0^{-1} \mathcal{J}_{I_c} \mathcal{L}_0^{-1} \pi_E } \nonumber \\[0.2cm]
&= 
\frac{\kappa}{2\Gamma}
\frac{\left(1 + \frac{1}{\bar{n}_h}\right)}{\left({1+\frac{1}{\bar{n}_{\rm ave}}}\right)}
\left[\biggl(1 + \frac{\kappa}{\gamma_{\rm cl}}\biggl)^{\!-1}
+
\frac{\left(1 + \frac{1}{\bar{n}_c}\right)}{\left({1+\frac{1}{\bar{n}_{\rm ave}}}\right)}
\biggl(1 + \frac{\gamma_{\rm cl}}{\kappa}\biggl)^{\!-1}\right],
\\[0.3cm]
p_3^{} &= \tr{ \mathcal{J}_{E_h} \mathcal{L}_0^{-1} \mathcal{J}_{I_h} \mathcal{L}_0^{-1} \pi_E } \nonumber \\[0.2cm]
&=%
\frac{\kappa}{2\Gamma}
\frac{\bar{n}_h \gamma_h}{\bar{n}_c \gamma_c}
\frac{\left(1 + \frac{1}{\bar{n}_h}\right)}{\left({1+\frac{1}{\bar{n}_{\rm ave}}}\right)}
\left[
\biggl(1 + \frac{\kappa}{\gamma_{\rm cl}}\biggl)^{\!-1}
+
\frac{\left(1 + \frac{1}{\bar{n}_h}\right)}{\left({1+\frac{1}{\bar{n}_{\rm ave}}}\right)}
\biggl(1 + \frac{\gamma_{\rm cl}}{\kappa}\biggl)^{\!-1}
\right],
\\[0.3cm]
p_4^{} &= \tr{ \mathcal{J}_{E_c} \mathcal{L}_0^{-1} \mathcal{J}_{I_c} \mathcal{L}_0^{-1} \pi_E } \nonumber \\[0.2cm]
&= 
\frac{\kappa}{2\Gamma}
\frac{\bar{n}_c \gamma_c}{\bar{n}_h \gamma_h}
\frac{\left(1 + \frac{1}{\bar{n}_c}\right)}{\left({1+\frac{1}{\bar{n}_{\rm ave}}}\right)}
\left[
\biggl(1 + \frac{\kappa}{\gamma_{\rm cl}}\biggl)^{\!-1}
+
\frac{\left(1 + \frac{1}{\bar{n}_c}\right)}{\left({1+\frac{1}{\bar{n}_{\rm ave}}}\right)}
\biggl(1 + \frac{\gamma_{\rm cl}}{\kappa}\biggl)^{\!-1}
\right].
\end{align}
It is straightforward to verify that $p_1 + p_2 + p_3 + p_4 = 1$.
Remarkably,
\begin{subequations}
\label{maser_probs}
\begin{align}
p_1 - p_2 &= \frac{1}{2\Gamma}\biggl(\frac{1}{\kappa} + \frac{1}{\gamma_{\rm cl}}\biggr)^{\!-1}\biggl(1 + \frac{1}{\bar{n}_{\rm ave}} \biggr)^{\!-1}\biggl(\frac{1}{\bar{n}_c} - \frac{1}{\bar{n}_h}\biggr), \\[0.2cm]
\frac{p_3}{p_1} = \frac{p_2}{p_4} &= \frac{(\bar{n}_h + 1) \gamma_h}{(\bar{n}_c + 1) \gamma_c}.
\label{maser_prob_ratio}
\end{align}
\end{subequations}
As expected from Eq.~\eqref{prob_curr}, $p_1 - p_2$ relates to the particle current. In this model, this is indeed evidenced by its dependence on the difference in Bose occupations, $\bar{n}_h - \bar{n}_c$.
These expressions are plotted in Fig.~\ref{fig:maser_stats}\,(b) of the main text, where we also set $\gamma_h = \gamma_c$ leading to the following conclusions:
\begin{itemize}
    \item In the refrigeration regime ($\bar{n}_h < \bar{n}_c$), we have $p_3 < p_1 < p_2 < p_4$. 
    \item In the engine regime ($\bar{n}_h > \bar{n}_c$), we have  $p_4 < p_2 < p_1 < p_3$.
\end{itemize}
Thus, the idle cycle probabilities bound those of useful cycles across all parameter ranges.

\subsection{Useful cycle probabilities}
The probability of observing a useful cycle is  $p_u := p_1 + p_2$. 
Because the post-injection subspace in this case is one-dimensional, it turns out that the cycles are independent of each other. 
As a consequence, the probability of observing $n$ idle cycles between two useful ones [Eq.~\eqref{pusen}] will be given simply by the geometric distribution 
\begin{equation}
    \mathbb{P}_{\rm u}(n) = p_u(1-p_u)^n.
\end{equation}

\begin{figure}[t]
    \centering
    \includegraphics[width=\textwidth]{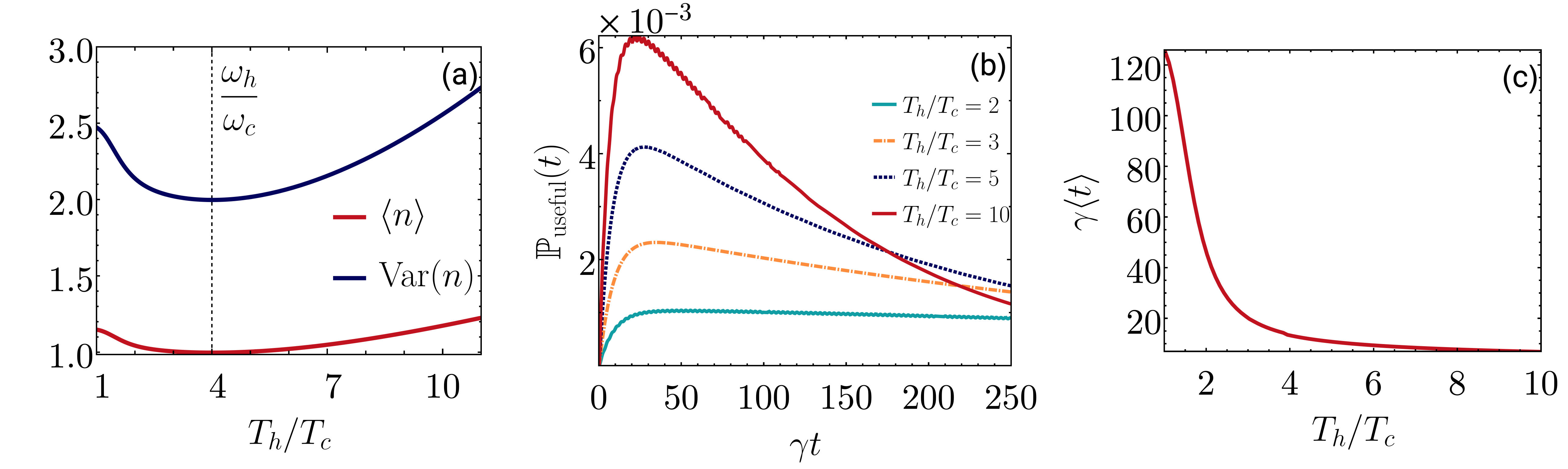}
    \caption{Probabilities of useful cycles as a function of {\bf (a)} number of intervening idle cycles $n$ [Eq.~\eqref{pusen}] and {\bf (b)} time interval $t$ between them [Eq.~\eqref{puset}]. A steeper temperature gradient between the baths prompts faster heat transport while also encouraging more idle cycles.
    {\bf (c)} The average time between two useful cycles as a function of bath temperature ratio. This curve resembles the average cycle times from Fig.~\ref{fig:maser_stats}\,(c).
    }
    \label{fig:pu}
\end{figure}

Figure~\ref{fig:pu}\,(a) shows the mean and variance: 
\begin{equation}
\label{maser_mean}
\langle n \rangle = \frac{1}{p_u} - 1 = \frac{p_{\rm id}}{p_u}, \qquad {\rm Var}(n) = \frac{1-p_u}{p_u^2},
\end{equation}
of this geometric distribution.
We find the optimal temperature ratio $T_h/T_c = \omega_h/\omega_c$ minimizes the mean number of idle cycles.
However, a steeper temperature gradient between the baths does not necessarily result in efficient heat transport, as it also promotes the participation of idle cycles in system dynamics.
To quantify this, we turn to the time between two useful cycles given in Eq.~\eqref{puset} and computed numerically in this case.
The results plotted in Fig.~\ref{fig:pu}\,(b) show that a steeper temperature gradient facilitates faster heat transport.
Finally, as a measure of occurrence of useful cycles, we plotted the average time between two useful cycles in Fig.~\ref{fig:pu}\,(c), obtained from $\gamma \langle t \rangle = \gamma \int_0^\infty dt~t\,\mathbb{P}_{\rm u}(t)$ derived via Eq.~\eqref{puset}.
As the bath temperature ratio increases, the average time descends in a manner similar to the average cycle times as seen in Fig.~\ref{fig:maser_stats}\,(c) in main text.

\subsection{Intermittency as average number of cycles}

The mean number of intervening idle cycles for this model is given in Eq.~\eqref{maser_mean}: $\langle n \rangle = p_{\rm id}/p_u$. We claim that when relaxation rates of both baths are equal ($\gamma_h = \gamma_c$), the idle cycles always dominate the useful ones, $p_{\rm id} \geq p_{u}$, implying that $\langle n \rangle \geq 1$ for all parameter ranges.

A simple proof is to determine the sign of $p_{\rm id} - p_u$ at $\gamma_h = \gamma_c$. Using the relations in Eq.~\eqref{maser_probs},
\begin{equation}
\begin{split}
p_{\rm id} - p_u &= p_3 + p_4 - p_1 - p_2 \\
&= p_1 \left(\frac{\bar{n}_h + 1}{\bar{n}_c + 1} - 1\right) + p_2 \left(\frac{\bar{n}_c + 1}{\bar{n}_h + 1} - 1\right) \\
&=(\bar{n}_h - \bar{n}_c)  \left(\frac{p_1}{\bar{n}_c + 1}- \frac{p_2}{\bar{n}_h + 1}\right)\geq 0,
\end{split}
\end{equation}
for all ranges of parameters.
The above inequality becomes apparent by noting that $\bar{n}_h \geq \bar{n}_c$ implies $p_1 \geq p_2$ and vice-versa.

\end{document}